\documentclass[aoas]{imsart}

\RequirePackage{graphicx}
\RequirePackage{pdflscape}
\RequirePackage{booktabs,multirow,siunitx,algorithm,algpseudocode}
\RequirePackage{amsthm,amsmath,amsfonts,amssymb,mathtools,bm}
\RequirePackage[authoryear]{natbib}
\RequirePackage[colorlinks,citecolor=blue,urlcolor=blue,linkcolor=blue]{hyperref}
\RequirePackage[nameinlink,capitalize]{cleveref}

\startlocaldefs
\usepackage{rotating}
\usepackage{float}
\usepackage{tikz}
\usetikzlibrary{arrows.meta, positioning, calc}

\usepackage{multirow}
\usepackage{url}

\theoremstyle{plain}

\theoremstyle{definition}

\def\hat{ \widehat}

\def\ba{\mbox{\boldmath $a$}}

\def\bu{\mbox{\boldmath $u$}}

\def\balpha{\mbox{\boldmath $\alpha$}}
\def\bbeta{\mbox{\boldmath $\beta$}}

\def\bepsilon{\mbox{\boldmath $\epsilon$}}

\def\bmu{\mbox{\boldmath $\mu$}}

\def\btheta{\mbox{\boldmath $\theta$}}

\def\1{\mbox{\boldmath $1$}}

\def\0{\mbox{\boldmath $0$}}

\def\be{\mbox{\boldmath $e$}}

\def\bO{\mbox{\bf O}}

\def\bx{\mbox{\boldmath $x$}}

\def\by{\mbox{\boldmath $y$}}
\def\bz{\mbox{\boldmath $z$}}

\endlocaldefs

\begin{document}

\newcommand{\multic}[1]{\multicolumn{2}{c}{#1}}
\setlength{\tabcolsep}{-0.5pt}
\def\phm{\phantom{-}}
\def\piii{{\footnotesize  \phantom{***}}}
\def\pii {{\footnotesize  \phantom{** }}}
\def\pstar{{\footnotesize \phantom{*  }}}
\def\phdot{\phantom{.}}
\def\sss{{\tiny $\!\!\!*\!\!*\!\!*\!\!$}}

\begin{frontmatter}

\title{Multivariate Spatio-Temporal Regression with\\
Penalized Model Selection and\\ an Empirical Application}
\runtitle{Multivariate spatio-temporal regression with penalized model selection}

\begin{aug}
\author[A]{\fnms{Ryuei} \snm{Nishii}\ead[label=e1]{nishii.ryuei@nagasaki-u.ac.jp}}
\author[B]{\fnms{Saeko} \snm{Ohta}\ead[label=e2]{s.ota@meio-u.ac.jp}}
\author[C]{\fnms{Shojiro} \snm{Tanaka}\ead[label=e3]{sh-tanaka@hue.ac.jp}}

\address[A]{School of Information and Data Sciences, Nagasaki University, Nagasaki 852-8521, Japan \printead[presep={,\ }]{e1}}
\address[B]{Faculty of Human Health Sciences, Meio University, Nago, Okinawa 905-8585, Japan \printead[presep={,\ }]{e2}}
\address[C]{Faculty of Media Business, Hiroshima University of Economics, Hiroshima 731-0192, Japan \printead[presep={,\ }]{e3}}
\end{aug}

\begin{abstract}
This paper develops the statistical foundations of a multivariate general nesting spatio-temporal (MGNST) regression framework 
for analyzing spatial, temporal, and cross-equation dependence among responses.
Four parameter matrices represent spatial lag dependence, spatial error dependence,
temporal autoregression, and contemporaneous error covariance.
Their off-diagonal elements allow dependence to propagate within and across responses.
Matrix restrictions yield eleven specifications encompassing multivariate spatial
autoregressive models, multivariate spatial error models, vector autoregressive
models with exogenous variables, and independent spatio-temporal regressions as
special cases.

We establish identifiability conditions using instrumental-variable rank conditions
and introduce penalized likelihood estimation and information criteria based on
effective degrees of freedom.
Monte Carlo experiments examine three data-generating models, three sample sizes,
and three levels of spatial dependence.
Correct-selection rates under penalized AIC generally increase with sample size and
the strength of spatial dependence, while parameter recovery improves as the sample
size increases.

For socioeconomic data from 198 municipalities in Japan's Kansai region, penalized AIC selects the full MGNST model, 
whereas penalized BIC selects a response-wise independent spatial error model.
Despite selecting models of different complexity, both criteria support temporal persistence and spatial error dependence.
The pAIC-selected MGNST model reduces strong spatial autocorrelation in the responses to negligible residual levels, 
demonstrating its usefulness for identifying and comparing multivariate spatio-temporal dependence structures.
\end{abstract}


\end{frontmatter}

\section{Introduction}        
Spatial and spatio-temporal regression models play an important role in regional science and econometrics, 
as well as in environmental studies, epidemiology, and public policy analysis.
In many applications, multiple regional outcomes are observed simultaneously and evolve through interactions across space and time.
Examples include demographic composition, industrial structure, income, housing markets,
and transportation demand \citep{anselin1988,lesage2009}, 
as well as environmental indicators and disease incidence \citep{cressie2011,elliott2004}.
Such outcomes may exhibit within-response spatial dependence, temporal persistence, cross-response interactions, 
and contemporaneous dependence among unobserved disturbances.
Analyzing each response separately may therefore overlook important features of the underlying regional system.

A wide range of spatial regression models has been developed to represent spatial dependence.
The spatial autoregressive (SAR) model describes dependence transmitted through neighboring responses,
whereas the spatial error model (SEM) represents spatial correlation in the disturbance process \citep{elhorst2014}.
Combining these two forms of dependence yields the spatial autoregressive model with autoregressive disturbances (SARAR), 
which includes both spatially lagged responses and spatially correlated errors.
A more general specification is the general nesting spatial (GNS) model,
which additionally incorporates spatially lagged explanatory variables \citep{anselin2003,burridge2017,ward2019}.
Spatial models have also been extended to simultaneous systems with spatially interrelated equations \citep{kelejian2004}.

Multivariate extensions include multivariate spatial autoregressive models estimated by quasi-maximum likelihood \citep{yang2017}, 
multivariate spatial error formulations, and hierarchical Bayesian or latent Gaussian process models \citep{banerjee2014,mastrantonio2019,bradley2018}.
Temporal dependence is commonly represented by vector autoregressive structures, 
which allow both persistence within each response and lagged interactions across responses \citep{lutkepohl2005,kilian2017}.
Nevertheless, likelihood-based multivariate spatial models have primarily emphasized spatial lag dependence.
A unified regression framework that simultaneously accommodates spatial lag dependence, spatial error dependence, 
temporal autoregressive effects, and cross-equation error covariance has received less attention.

This paper develops a multivariate general nesting spatio-temporal (MGNST) regression model to address this problem.
For $K$ response variables, the model represents spatial lag dependence by a matrix $R$, spatial dependence in the error process by a matrix $\Lambda$,
temporal autoregressive dependence by a matrix $A$, and contemporaneous cross-equation error covariance by a matrix $\Sigma$.
The off-diagonal elements of these matrices permit dependence to propagate not only within a response but also across different responses.
The formulation thus extends the classical GNS model to multivariate spatio-temporal settings and includes multivariate SAR, multivariate SEM, 
vector autoregressive models with exogenous variables (VARX), and independent spatio-temporal regressions as special cases.

A bivariate simultaneous-equation model incorporating spatial lag dependence, temporal autoregressive dependence, and cross-equation error covariance, 
but not spatial error dependence, was first introduced by \citet{nishii2025}.
A related application-oriented implementation of the MGNST framework is described by \citet{ohta2026b} 
as part of an open-source geospatial workflow centered on an interactive web-mapping application.
The conference paper uses conventional maximum likelihood estimation and compares models using AIC and BIC, 
without addressing parameter identifiability or penalized estimation.
This study develops the statistical methodology in detail by establishing sufficient identifiability conditions, 
introducing penalized likelihood estimation and effective-degrees-of-freedom-based information criteria, 
and conducting an extensive Monte Carlo evaluation.

The framework is distinguished by three main features. 
First, simple  full, diagonal, or zero restrictions on $(R,\Lambda,A,\Sigma)$ generate eleven 
candidate models ranging from the full MGNST specification to independent univariate regressions. 
These models can therefore be estimated and compared  within a common likelihood-based framework. 
In contrast to the multivariate SAR model of \citet{yang2017}, the proposed formulation jointly allows for 
spatial dependence in both the response and error processes, together with temporal autoregressive effects 
and cross-equation error covariance.

Second, although identification has been studied for spatial econometric models \citep{lee2016}, 
we establish explicit sufficient conditions for the identifiability of the structural spatial parameters in the proposed
multivariate framework.
Simultaneously including spatial lag and spatial error terms can make different dependence mechanisms difficult to distinguish, 
particularly when their induced regressors are nearly linearly dependent. 
For the proposed MGNST model, we derive sufficient identifiability conditions using instrumental-variable rank conditions. 
The conditions clarify the roles of the explanatory-variable design, the spatial weight matrix, 
and the spatially transformed instruments in distinguishing the spatial lag and spatial error parameters.

Third, we develop a penalized likelihood procedure to stabilize estimation of the spatial dependence matrices $R$ and $\Lambda$
 \citep{konishi1996,hastie2009}.
Because the number of structural parameters increases rapidly with the number of responses, 
direct maximum likelihood estimation may be unstable, 
especially when the spatial signal is weak or the sample size is limited. 
The proposed quadratic penalty regularizes the estimation of the spatial parameters 
while retaining the likelihood-based structure of the model. 
Effective degrees of freedom derived from the penalized Hessian are then used to construct penalized AIC and BIC criteria, 
denoted by pAIC and pBIC. 
These criteria permit model complexity to be evaluated in terms of the effective, 
rather than merely nominal, number of estimated parameters.

Monte Carlo experiments show that model-selection and parameter-estimation performance generally 
improve with sample size and spatial-signal strength,
although closely related nested models remain difficult to distinguish under weak spatial dependence.

The empirical analysis uses municipal-level socioeconomic data from the Kansai region of Japan.
The two responses exhibit strong positive spatial autocorrelation. 
Among the eleven candidate specifications, pAIC selects the full MGNST model, 
whereas pBIC favors a simpler response-wise independent spatial error model. 
Despite this difference, both the criteria support temporal persistence and spatial error dependence. 
Estimated innovations from the pAIC-selected model exhibit no significant residual spatial autocorrelation, 
indicating that the proposed framework effectively accounts for the observed spatial dependence.

The remainder of this paper is organized as follows. 
Section~2 reviews the spatial regression models forming the basis of the proposed framework. 
Section~3 introduces the MGNST model, which presents the eleven candidate specifications, 
and establishes sufficient identifiability conditions. 
Section~4 develops likelihood-based estimation, penalized inference, and model selection 
using effective degrees of freedom. 
Section~5 reports the simulation experiments and the empirical analysis of the Kansai municipal data. 
Section~6 summarizes the main findings and discusses extensions to multiple temporal transitions and higher-order temporal dependence.


\section{Background on Spatial Regression Models}           
Let $\by = (y_1,\ldots,y_n)'$ be a response vector observed over $n$ regions,
let $W$ be an $n\times n$ row-standardized spatial weight matrix, and
let $X=[\bx_1,\ldots,\bx_p]$ be an $n\times p$ design matrix.
The columns of $X$ may include both observed explanatory variables and their spatially lagged counterparts; 
that is, for an explanatory variable $\bx$, 
$X$ may contain $W\bx$ representing the weighted average of $\bx$ over neighboring regions.
A widely used framework in spatial econometrics is the general nesting spatial (GNS) model,
\begin{equation}
 \by = \rho W\by + X\bbeta + \bu, \qquad
 \bu = \lambda W\bu + \bepsilon, \qquad
 \bepsilon \sim N(\0, \, \sigma^2I_n),
\end{equation}
where $\rho$ and $\lambda$ denote the spatial lag and spatial error parameters, respectively.
Thus, the GNS model can jointly accommodate spatially lagged responses, 
spatially lagged explanatory variables, and spatial dependence in the disturbances.
The spatial autoregressive (SAR) model and the spatial error model (SEM) are obtained 
as special cases by setting $\lambda=0$ and $\rho=0$, respectively,
when $X$ contains no spatially lagged explanatory variables.

For longitudinal regional data, temporal dependence is commonly incorporated through an autoregressive term. 
A spatio-temporal extension of the GNS model is given by
\begin{equation}
 \by_t = \rho W\by_t + X_t \, \bbeta + \alpha \, \by_{t-1} + \bu_t, \quad
 \bu_t = \lambda W\bu_t + \bepsilon_t, \quad
 \bepsilon_t \sim N(\0, \, \sigma^2I_n),
\end{equation}
where $\alpha$ is the temporal autoregressive coefficient.

These models have been used to analyze diverse environmental and socioeconomic phenomena, 
such as urban ozone concentrations \citep{huerta2004} and regional unemployment dynamics \citep{halleckvega2016}.
Most existing formulations, however, focus on a single outcome 
and do not directly accommodate dependence across multiple response processes.
This limitation motivates a multivariate formulation for jointly analyzing interrelated outcomes observed over space and time.


\section{Multivariate Spatio-Temporal Regression} 
The MGNST framework proposed here extends the GNS formulation 
by jointly incorporating spatial lag dependence, spatial error dependence, temporal autoregressive effects, 
and cross-equation covariance structures within a unified multivariate model.

\subsection{Multivariate General Nesting Spatio-Temporal Models}  
Consider a multivariate spatio-temporal regression setting with $K$ response vectors 
$\by_{1,t}, \ldots, \by_{K,t}$ observed over $n$ regions at time $t$.
Define the stacked response vector and coefficient matrices by
\begin{align}
 \by_t =
 \begin{bmatrix}
  \by_{1,t} \\
  \vdots    \\
  \by_{K,t}
 \end{bmatrix},
 \ \  
 R =
 \begin{bmatrix}
  \rho_{11} & \cdots & \rho_{1K} \\
  \vdots    & \ddots & \vdots    \\
  \rho_{K1} & \cdots & \rho_{KK}
 \end{bmatrix},
 \ \  
 A =
 \begin{bmatrix}
  \alpha_{11} & \cdots & \alpha_{1K} \\
  \vdots      & \ddots & \vdots      \\
  \alpha_{K1} & \cdots & \alpha_{KK}
 \end{bmatrix},
 \ \  
 \Lambda =
 \begin{bmatrix}
  \lambda_{11} & \cdots & \lambda_{1K} \\
  \vdots       & \ddots & \vdots       \\
  \lambda_{K1} & \cdots & \lambda_{KK}
 \end{bmatrix}.
\end{align}

Here, the matrices $R$, $A$, and $\Lambda$ respectively represent spatial lag dependence,
temporal autoregressive dependence, and spatial dependence in the error process.
We assume throughout that $I_{Kn} - R \otimes W$ and
$I_{Kn} - \Lambda \otimes W$ are non-singular.
Then, {\bf the MGNST regression model} is defined by
\begin{equation}
\by_t = (R \otimes W) \, \by_t + X_t^* \, \bbeta^* 
      + (A \otimes I_n) \, \by_{t-1} + \bu_t ,
\end{equation}
where $W : n \times n$ denotes a row-standardized spatial weight matrix,
and
\[ X_t^* =
\begin{bmatrix}
 X_{1,t}^* & \cdots & \bO \\
 \vdots    & \ddots & \vdots \\
 \bO       & \cdots & X_{K,t}^*
\end{bmatrix}
: Kn \times p^*
\]
is a block-diagonal design matrix with $X_{k,t}^* : n \times p_k^*$ for $k=1,\ldots,K$.
The regression coefficient vector is denoted by $\bbeta^* : p^* \times 1$,
where $p^* = p_1^* + \cdots + p_K^*$.

The error vector
$ \bu_t = \big( \bu_{1,t}', \ldots, \bu_{K,t}' \big)' : Kn \times 1 $
is assumed to follow the spatial error structure
\begin{equation}
 \bu_t = (\Lambda \otimes W) \, \bu_t + \bepsilon_t, \quad \bepsilon_t \sim N(\0_{Kn},\ \Sigma \otimes I_n),
\end{equation}
where
$\Sigma = \big(\ \sigma_{k\ell} \ \big) : K \times K$
is an unknown cross-equation covariance matrix.

The three terms on the right-hand side of Eq.~(4) represent spatial spillover,
exogenous regression, and temporal AR(1) effects, respectively. 
Together with the spatial error structure in Eq.~(5), the model includes multivariate SAR,
multivariate SEM, VARX, and response-wise independent spatio-temporal
regressions as special cases. In particular, VARX is obtained by setting $R = \Lambda = \bO$.

Although the design matrices $X_{1,t}^*, \ \ldots, \ X_{K,t}^*$ may share common explanatory variables,
their regression coefficients are allowed to differ across response variables.
This flexibility is important in regional applications 
where common socioeconomic factors may affect multiple responses in different ways.

We next introduce the standard form of the MGNST model. Let 
 $ Y_{t-1} = [\by_{1,t-1},\ldots,\by_{K,t-1}] : n\times K $
denote the matrix of observed response vectors at time $t-1$, and let 
 $ \balpha_k=(\alpha_{k1},\ldots,\alpha_{kK})' $
denote the AR(1) coefficient vector for the $k$-th response, $k=1,\ldots,K$.
Then, the second and third terms on the right-hand side of Eq.~(4) are simply combined as 
\begin{equation}
\begin{bmatrix}
  X_{1,t}^*  &  Y_{t-1} & \cdots & \bO        & \bO     \\
  \vdots     & \ddots   & \ddots & \vdots     & \vdots  \\
  \vdots     & \vdots   & \ddots & \ddots     & \vdots  \\
  \bO        & \bO      & \cdots & X_{K, t}^* & Y_{t-1} \\
\end{bmatrix}
\begin{bmatrix}
 \bbeta_1^*   \\[-2mm]
 \balpha_{1}  \\[-1mm]
 \vdots       \\[-2mm]
 \bbeta_K^*   \\[-1mm]
 \balpha_{K}  \\[-1mm]
\end{bmatrix} 
 \ = \ 
\begin{bmatrix}
  X_{1,t} & \cdots & \cdots & \bO  \\
  \vdots  & \ddots & \ddots & \vdots\\
  \vdots  & \ddots & \ddots & \vdots\\
  \bO     & \cdots & \cdots & X_{K,t} \\
\end{bmatrix}
  \bbeta \ = \ X_t \; \bbeta
\end{equation}
\noindent
where $X_{k, t} = [X_{k, t}^*, \ Y_{t-1} ] : n \times p_k\ (p_k := p_k^* + K)$ are composite design matrices 
with $\bbeta_k = (\bbeta_k^{*'}, \, \balpha_k')': p_k \times 1$ 
and $\bbeta := (\bbeta_1', ... , \bbeta_K')': p \times 1$ is a combined coefficient vector with $p := p_1 + \cdots + p_K$.
Then, Eqs.~(4)--(6) are unified into {\bf the standard form of the MGNST model} as
\begin{align}
 \by_t
 &=
 (R \otimes W)\by_t
 + X_t\bbeta
 + (I_{Kn}-\Lambda\otimes W)^{-1}\bepsilon_t,
 \quad
 \bepsilon_t
 \sim
 N(\0_{Kn},\Sigma\otimes I_n).
\label{eq:6.2}
\end{align}

\indent
In Eq.~(7), the design matrix $X_t$ includes the lagged response vector $\by_{t-1}$ as temporal regressors, 
together with the explanatory variables.
The MGNST model considered in this paper is therefore formulated as a conditional regression model
for $\by_t$ given $(\by_{t-1},X_t)$, and inference is based on the conditional density
$f(\by_t \mid \by_{t-1}, X_t) = f(\by_t \mid X_t)$ because $X_t$ includes $\by_{t-1}$.
Accordingly, the proposed estimation procedure does not require specification of the marginal distribution of the initial response vector 
or of the joint distribution of the entire temporal process.

The conditional expectation of $\by_t$ is given by
$ E(\by_t \mid X_t) = (I_{Kn} - R\otimes W)^{-1} \; X_t \, \bbeta, $
showing that the proposed MGNST model represents spatial spillover effects through the spatial lag component.

\subsection{Identifiability of the MGNST model parameters}   
The MGNST model jointly incorporates the spatial dependence matrices $R$ and $\Lambda$. 
Consequently, the identifiability of the structural dependence parameters becomes a fundamental issue for statistical inference. 

Let
\begin{align}
 S_R &= I_{Kn} - R \otimes W : Kn\times Kn, \qquad 
 S_\Lambda = I_{Kn} - \Lambda\otimes W : Kn\times Kn.
\end{align}
We assume that $S_R$ and $S_\Lambda$ are {\bf real matrices with positive determinants}.
Since the temporal index $t$ is not essential for the argument, it is omitted here.

Then, the MGNST model can be written as 
\begin{align}
 {S_R} \; \by = X\, \bbeta + \bu, \ \ \mbox{where} \ \ \bu = S_\Lambda^{-1} \, \bepsilon, \ \ 
 E ( \; \bepsilon \mid X ) = \0 \ \ \mbox{and} \ \ 
 \operatorname{Cov}(\bepsilon) = \Sigma \otimes I_n.
 \label{eqn:s_r}
\end{align}
Let
$ \btheta_0 = (R_0, \Lambda_0, \bbeta_0, \Sigma_0) $ be the true parameter, and
$ \btheta   = (R,   \Lambda,   \bbeta,   \Sigma  ) $ be a candidate parameter. 
Then, we have the following two propositions for identifiability.
\\[-1mm]

\noindent
{\bf Proposition 1}: (Identifiability of $R$ and $\bbeta$)
{
Let $\bmu_0 = ( \bmu_{1,0}', \ ..., \ \bmu_{K,0}' )' = E(\by): Kn \times 1$ be the true mean vector, and let
\begin{align}
  G_{\mu_0} 
&=
 \begin{pmatrix}
  W \bmu_{1,0} & \cdots & W \bmu_{K,0} & \cdots & \0           & \cdots       & \0           \\
  \vdots       & \ddots & \vdots       & \ddots & \vdots       & \ddots       & \vdots       \\
  \0           & \cdots & \0           & \cdots & W \bmu_{1,0} & \cdots       & W \bmu_{K,0} \\
 \end{pmatrix}
 : Kn \times K^2.
\end{align}
If $ \operatorname{rank} \left[ \ G_{\mu_0}, \ X \ \right] = K^2 + p $ (full column rank), 
then $R$ and $\bbeta$ are identifiable.

This is a population-level sufficient condition.
\\[-2mm]


\noindent
{\bf Proposition 2}: (Identifiability of $\Lambda$ and $\Sigma$)

Suppose that $R$ and $\bbeta$ have already been identified.
If the matrices $I_n$, $W$, $W'$, and $W'W$ are linearly independent over the real numbers, 
then $\Lambda$ and $\Sigma$ are identifiable.

This assumption is commonly satisfied in practical applications unless the spatial configuration is degenerate 
because $W$ is generally nonsymmetric after row standardization and 
therefore $W$, $W'$, and $W'W$ rarely satisfy nontrivial linear relations.
The condition may fail for symmetric or otherwise degenerate spatial configurations, 
but it is typically satisfied for irregular row-standardized spatial weight matrices.
\\[-2mm]

Proofs of both propositions are given in Appendix~A. 
Note that the rank condition in Proposition~1 and the linear-independence condition in Proposition~2 are sufficient 
but not necessary conditions for identifiability.

\section{Estimation and Model Selection}    
\subsection{Likelihood equations for the MGNST models}        
The log likelihood of $\by_t$ defined in Eq.~(7) is given by
\begin{align}
\ell(R,\Lambda,\bbeta,\Sigma)
   &= -\tfrac{Kn}{2}\log(2\pi)
     + \log \, \big| S_R       \big|
     + \log \, \big| S_\Lambda \big|
     - \tfrac{n}{2}\log| \Sigma |
     - \tfrac{1}{2}Q 
\end{align}
\noindent
where $S_\Lambda, \ S_R$ are defined by Eq.~(8), and
\begin{align}
 Q &= \bz'\, (\Sigma^{-1}\otimes I_n) \, \bz \ \ \mbox{with the residual vector:} \ \ 
 \bz = S_\Lambda \big\{ S_R \; \by_t - X_t \, \bbeta \big\}.
\label{eq:6.4}
\end{align}
If $R$ and $\Lambda$ are both known matrices, the MGNST model is just a multivariate regression model with correlated errors.

Based on the matrix derivative identity : $\partial \, \log \, | \Sigma^{-1} | / \partial \, \Sigma^{-1} = \Sigma$,
likelihood equations with respect to $\bbeta$ and $\Sigma^{-1} = \big(\ \sigma^{k, \ell} \ \big)$ are given by
\begin{align}
     \partial\, \ell / \, \partial \bbeta' \ 
 &= \ X_t' \, S_{\Lambda}' \, ( \Sigma^{-1} \otimes I_n) \, \bz \ = \ \0
 \\[1mm]
     \partial\, \ell / \, \partial \sigma^{k,\ell} \ 
 &= \ \frac{n}{2} \sigma_{k, \ell}
  -    \tfrac12 \, \bz' \left\{ \mathbf{e}_k \mathbf{e}_\ell' \otimes I_n \right\}
                        \bz 
\ = \ \frac{n}{2} \sigma_{k, \ell} - \frac{1}{2} \bz_k' \bz_\ell \ = \ 0, \ \ \mbox{where}
 \\ \notag
 \be_k \ &= \ ( \ \underbrace{0, \dots, 0}_{k-1}, \; 1 \; , \underbrace{0, \dots, 0}_{K-k} \ )' : K \times 1.
\end{align}
Here, $\bz_k \ (k = 1,...,K)$ are $n$-dimensional subvectors of the residuals $\bz = \big(\ \bz_1', \ ... , \ \bz_K' \ \big)'$
 defined by Eq.~(12).
Thus, we have
\begin{align}
 \hat{\bbeta} &= \hat{\bbeta}(R, \Lambda, \Sigma) \ = \ B \, \by_t,  \ \ \ \mbox{where}
 \\
 B            &= \left\{ X_t' \, S_\Lambda' \left( \Sigma^{-1} \otimes I_n \right) S_\Lambda X_t \, \right\}^{-1} 
                         X_t' \, S_\Lambda' \left( \Sigma^{-1} \otimes I_n \right) S_\Lambda \, S_R \\
 \hat{\Sigma} &= \hat{\Sigma}(R, \Lambda, {\bbeta}) 
             \ = \ \frac{1}{n}
                        \begin{pmatrix}
                         \bz_{1}' \, \bz_{1} & \cdots & \bz_{1}' \, \bz_{K} \\
                         \vdots              & \ddots & \vdots              \\
                         \bz_{K}' \, \bz_{1} & \cdots & \bz_{K}' \, \bz_{K} \\
                        \end{pmatrix}.
\end{align}
\noindent
The covariance matrix of $\hat{\bbeta}$ is
\begin{align}
 \Psi &= B \; \mbox{Cov}(\by) \, B' 
       = \left\{ X_t' \, S_\Lambda' \left( \Sigma^{-1} \otimes I_n \right) S_\Lambda \, X_t \right\}^{-1}.
 \label{37}
\end{align}
Equation~(18) gives the conditional covariance matrix of $\hat{\bbeta}$, 
treating the estimated spatial dependence matrices $R$ and $\Lambda$ as fixed. 
Because it does not account for the additional uncertainty arising from the estimation of $(R,\Lambda)$, 
inference based solely on $\Psi$ is approximate and may overstate the statistical significance of the regression coefficients. 


\subsection{Estimation procedure for the MGNST models}       
Let us partition the design matrix as 
$X_t = \operatorname{diag} \big[\, X_{1,t} , \ ... , \ X_{K,t} \, \big]$ 
with $X_{k,t}: n \times p_k$ in Eq.~(6).
The estimation procedure is as follows.
\begin{enumerate}
 \item[S1] Obtain initial parameter estimates from separate GNS models.
   \begin{enumerate}
     \item For each $k = 1,...,K$, estimate $\rho_{kk}$, $\lambda_{kk}$ and $\sigma_{kk}$ of the GNS given by \\[-7mm]
           \begin{align}
             \by_{k, t}= \rho_{kk} W \by_{k, t}
                       + X_{k,t} \bbeta_k
                       + (I - \lambda_{kk} W)^{-1} \bepsilon_{k, t} 
              \ \ \mbox{with} \ \ \bepsilon_{kt} \sim N(\0_n, \sigma_{kk} I) \notag
           \end{align}
     \item Set \ \ 
             ${R} = \begin{pmatrix}
                         \hat{\rho}_{11} & \cdots & 0               \\
                         \vdots          & \ddots & \vdots          \\
                         0               & \cdots & \hat{\rho}_{KK} \\
                        \end{pmatrix}$, \ \ 
             ${\Lambda}
                 = \begin{pmatrix}
                     \hat{\lambda}_{11} & \cdots & 0                  \\
                     \vdots             & \ddots & \vdots             \\
                     0                  & \cdots & \hat{\lambda}_{KK} \\
                   \end{pmatrix}$, \ \ 
            $\hat{\Sigma}
                = \begin{pmatrix}
                   \hat{\sigma}_{11} & \cdots & 0                 \\
                   \vdots            & \ddots & \vdots            \\
                   0                 & \cdots & \hat{\sigma}_{KK} \\
                  \end{pmatrix} $,
               \\[1mm] 
         estimated by each GNS for $k=1,...,K$.
   \end{enumerate}
 \item[S2] Given $R$ and $\Lambda$, obtain the optimal values of $(\bbeta,\Sigma)$ as follows.
   \begin{enumerate}
     \item[(c)] Obtain $\hat{\bbeta} = \hat{\bbeta}(R, \Lambda, \hat{\Sigma})$ by Eq.~(15),
                and    $\hat{\Sigma} = \hat{\Sigma}(R, \Lambda, \hat{\bbeta})$ by Eq.~(17).
     \item[(d)] Repeat (c) until convergence.
   \end{enumerate}
 \item[S3] Obtain the maximized profile log-likelihood given $(R, \ \Lambda)$ using the preceding step: \\[-2mm]
       $$
            \ell(R, \Lambda)
          = \ell(R, \Lambda, \hat{\bbeta}, \hat{\Sigma})
          = - \frac{Kn}{2} \log (2 \pi e)  + \log \, | S_R | + \log \, | S_\Lambda |
            - \frac{ n}{2} \log \, | \hat{\Sigma}(R, \Lambda) |.
       $$ 
       Appendix~B gives a faster calculation of $| S_R |$ and $| S_\Lambda |$ based on the eigenvalues of $W$.
\item[S4] 
          Optimize $(R,\Lambda)$ over the admissible parameter space
          \[
          \mathcal{P}
          =
          \{ \, (R,\Lambda):
          |I_{Kn} - R\otimes W| > 0,\ 
          |I_{Kn} - \Lambda\otimes W| > 0 \, \}
          \]
          by applying the BFGS quasi-Newton method \citep{nocedal2006} 
          to maximize the penalized profile log-likelihood $\ell(R,\Lambda)$ defined in Section~4.4 and Appendix~C.
          The penalty
          $\gamma\operatorname{tr}(R'R+\Lambda'\Lambda)/2$
          regularizes the spatial dependence parameters, 
          while candidate values outside $\mathcal{P}$ are excluded so that the log-determinant terms remain real and well defined.
          If BFGS fails to converge, repeat the optimization using the Nelder--Mead method.
\end{enumerate}

Even within $\mathcal{P}$, estimation may be unstable when the profile likelihood is nearly flat, 
the spatial dependence components are strongly collinear, or the solution lies near the boundary of $\mathcal{P}$.
The penalty is therefore introduced to discourage excessively large spatial parameter estimates and stabilize the numerical estimation.


\subsection{Special cases of the MGNST model}   
The MGNST framework represents a wide class of linear regression models 
through suitable restrictions on the parameter matrices, including the MSAR and MSEM as special cases.

Figure~1 illustrates the hierarchical relationships among the proposed multivariate spatio-temporal models. 
The diagram summarizes spatial spillover effects, temporal autoregressive dependence, cross-variable interactions, 
and common explanatory effects represented in the MGNST framework. 
Each specification is labeled by a four-character model ID corresponding to $(R,\Lambda,A,\Sigma)$, 
where \texttt{1}, \texttt{d}, and \texttt{0} indicate a full matrix, 
a diagonal matrix, and the absence of the corresponding dependence structure, respectively. 
Thus, model \texttt{(1111)} denotes the unrestricted MGNST model, 
while model \texttt{(dddd)}, in which \texttt{d} indicates a diagonal parameter matrix, 
represents independent spatio-temporal regressions.
The number of free parameters for each specification is determined by $\bbeta$ 
and the restrictions imposed on $R$, $\Lambda$, $A$, and $\Sigma$.


\hspace*{-10mm}
 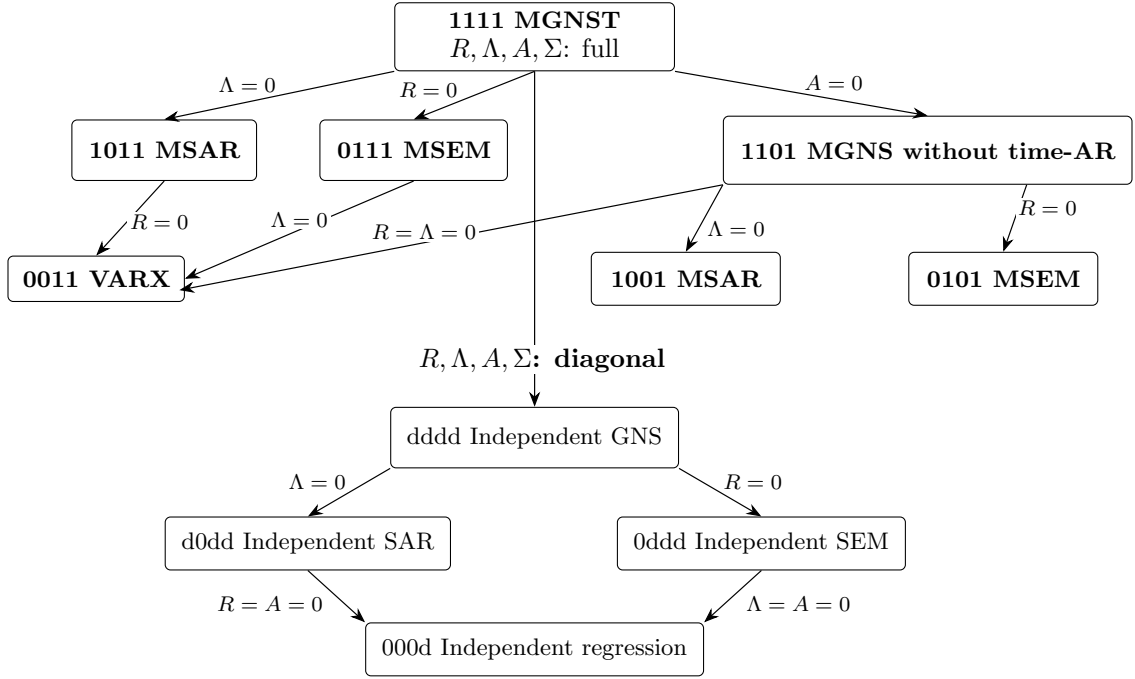
\begin{figure}[htbp]
\centering
\footnotesize
\begin{tikzpicture}[
  model/.style={
    draw,
    rounded corners=2pt,
    align=center,
    minimum width=37mm,
    minimum height=9mm,
    inner sep=3pt
  },
  midmodel/.style={
    draw,
    rounded corners=2pt,
    align=center,
    minimum width=22mm,
    minimum height=8mm,
    inner sep=3pt
  },
  smallmodel/.style={
    draw,
    rounded corners=2pt,
    align=center,
    minimum width=25mm,
    minimum height=7mm,
    inner sep=3pt
  },
  varxmodel/.style={
    draw,
    rounded corners=2pt,
    align=center,
    minimum width=22mm,
    minimum height=6mm,
    inner sep=2pt
  },
  arrow/.style={-{Stealth[length=2mm]}, line width=0.45pt},
  lab/.style={font=\scriptsize, fill=white, inner sep=1pt}
]

\node[model] (mgnst) at (0,0) {
  {\bf 1111 MGNST}\\
  {\small $R,\Lambda,A,\Sigma$: full}
};

\node[midmodel] (msar) at (-4.9,-1.5) {
 {\bf \ 1011 MSAR}
};

\node[midmodel] (msem) at (-1.6,-1.5) {
 {\bf \ 0111 MSEM}
};

\node[model] (mgns) at (5.2,-1.5) {
 {\bf \ 1101 MGNS without time-AR}
};

\node[varxmodel] (varx) at (-5.8,-3.2) {
 {\bf \ 0011 VARX}
};

\node[smallmodel] (msarNoA) at (2.0,-3.2) {
 {\bf \ 1001 MSAR}
};

\node[smallmodel] (msemNoA) at (6.2,-3.2) {
 {\bf \ 0101 MSEM}
};

\node[midmodel] (indgns) at (0,-5.3) {
  \ dddd Independent GNS
};

\node[smallmodel] (indsar) at (-3.0,-6.7) {
  \ d0dd Independent SAR
};

\node[smallmodel] (indsem) at (3.0,-6.7) {
  \ 0ddd Independent SEM
};

\node[smallmodel] (indreg) at (0,-8.1) {
  \ 000d Independent regression
};

\draw[arrow] (mgnst.south west) -- (msar.north)
  node[midway, above left, font=\scriptsize, fill=white, inner sep=1pt] {$\Lambda=0$};

\draw[arrow] (mgnst.south) -- (msem.north)
  node[pos=0.61, above left, font=\scriptsize, fill=white, inner sep=1pt] {$R=0$};

\draw[arrow] (mgnst.south east) -- (mgns.north)
  node[midway, above right, font=\scriptsize, fill=white, inner sep=1pt] {$A=0$};

\draw[arrow] (mgnst.south) -- (indgns.north)
  node[pos=0.86, lab] {\ \ \small {\bf $R,\Lambda,A,\Sigma$: diagonal}};

\draw[arrow] (msar.south) -- (varx.north)
  node[pos=0.55, right, font=\scriptsize, fill=white, inner sep=1pt] {$R=0$};

\draw[arrow] (msem.south) -- (varx.east)
  node[midway, above, font=\scriptsize, fill=white, inner sep=1pt] {$\Lambda=0$};

\draw[arrow] (mgns.south west) -- (-4.70,-3.35)
  node[pos=0.55, above, font=\scriptsize, fill=white, inner sep=1pt] {$R=\Lambda=0$};

\draw[arrow] (mgns.south west) -- (msarNoA.north)
  node[midway, below right, xshift=2mm, font=\scriptsize, fill=white, inner sep=1pt] {$\!\!\!\! \Lambda=0$};

\draw[arrow] ($(mgns.south east)+(-14mm,0)$) -- (msemNoA.north)
  node[midway, above right, font=\scriptsize, fill=white, inner sep=1pt] {$R=0$};

\draw[arrow] (indgns.south west) -- (indsar.north)
  node[midway, above left, font=\scriptsize, fill=white, inner sep=1pt] {$\Lambda=0$};

\draw[arrow] (indgns.south east) -- (indsem.north)
  node[midway, above right, font=\scriptsize, fill=white, inner sep=1pt] {$R=0$};

\draw[arrow] (indsar.south) -- (indreg.north west)
  node[pos=0.78, above left, xshift=-3.5mm, yshift=-0.5mm, font=\scriptsize, fill=white, inner sep=1pt] {$\ R=A=0$};

\draw[arrow] (indsem.south) -- (indreg.north east)
  node[pos=0.78, above right, xshift=3.5mm, yshift=-0.5mm, font=\scriptsize, fill=white, inner sep=1pt] {$\Lambda=A=0$};

\end{tikzpicture}
\caption{Hierarchical structure of the MGNST models and model IDs.
Models are obtained by imposing zero restrictions or diagonal restrictions on the component matrices
$R$, $\Lambda$, $A$, and $\Sigma$. \ \ 
The top three layers contain multivariate models with or without temporal autoregression, 
whereas the fourth and lower layers contain independent models with diagonal parameter matrices.}
\label{fig:diagram}
\end{figure}
 

\subsection{Model evaluation based on penalized likelihoods} 
\label{sec:model_evaluation}

We use ridge penalization to stabilize estimation of the spatial-lag and spatial-error parameters 
when these dependence structures are difficult to distinguish in finite samples. 
For candidate model $m$, write $\btheta_m = (\btheta_{1m}', \btheta_{2m}')'$, 
where $\btheta_{1m}$ contains the $q_{1m}$ free parameters in $R$ and $\Lambda$, 
and $\btheta_{2m}$ contains the $q_{2m}$ unpenalized regression, temporal, and covariance parameters. 
Let $D_m = \operatorname{diag}(I_{q_{1m}},0_{q_{2m}})$. 
For $\gamma \geq 0$, define
\begin{align}
   \hat{\btheta}_{m,\gamma}
 = \arg\max_{\theta_m \in \Theta_m}
   \left\{ \ell_m(\btheta_m) - \frac{\gamma}{2}\btheta_m' D_m \btheta_m \right\}.
 \label{eq:pen_estimator}
\end{align}
Thus, only the free spatial parameters are penalized; the penalty is used for stabilization, not parameter selection.

Let $\mathcal I_m(\hat{\btheta}_{m,\gamma})$ be the observed information matrix of the unpenalized log-likelihood evaluated 
at $\hat{\btheta}_{m,\gamma}$. 
The effective degrees of freedom are
\begin{align}
   d_{\mathrm{eff}}(m, \gamma)
 = \operatorname{tr}\! \left[ \, \mathcal I_m(\hat{\btheta}_{m,\gamma}) \{\mathcal I_m(\hat{\btheta}_{m,\gamma}) + \gamma D_m\}^{-1}
                       \right].
 \label{eq:effective_df}
\end{align}
This full-matrix expression allows dependence between the penalized and unpenalized parameter blocks. 
We then define 
 \[
  \mathrm{pAIC}(m,\gamma) = -2 \ell_m(\hat{\btheta}_{m,\gamma}) + 2        d_{\mathrm{eff}}(m,\gamma),
  \quad
  \mathrm{pBIC}(m,\gamma) = -2 \ell_m(\hat{\btheta}_{m,\gamma}) + \log(Kn) d_{\mathrm{eff}}(m,\gamma).
 \]
The unpenalized log-likelihood is evaluated at the penalized estimator. 
This adjustment follows generalized information criteria for penalized likelihood
\citep{konishi1996} and trace-based effective degrees of freedom for ridge estimators \citep{hastie2009}.

For either criterion $\mathrm{pIC}\in\{\mathrm{pAIC},\mathrm{pBIC}\}$, 
model and penalty selection are performed jointly as
$  (\hat m_{\mathrm{pIC}},\hat\gamma_{\mathrm{pIC}})
 = \arg\min_{m\in\mathcal M,\,\gamma \in \mathcal G} {\mathrm{pIC}(m,\gamma)} $,
where $\mathcal G$ is the prespecified search region. 
A coarse logarithmic grid locates the relevant region, followed by a refined one-dimensional search for each model. 
Equation~(\ref{eq:effective_df}) measures the local complexity of a fixed $(m,\gamma)$ and does not include search uncertainty 
over $\mathcal M\times\mathcal G$; pAIC and pBIC are therefore used for model comparison, not confirmatory inference.

For a prespecified $(m,\gamma)$, put
$\mathcal H_{p,m}=\mathcal I_m(\hat{\btheta}_{m,\gamma})+\gamma D_m$. 
A model-based first-order covariance approximation is
\begin{align}
   \widehat{\operatorname{Var}}( \hat{\btheta}_{m,\gamma} )
 = \mathcal H_{p,m}^{-1} \, \mathcal I_m(\hat{\btheta}_{m,\gamma}) \, \mathcal H_{p,m}^{-1}.
 \label{eq:pen_covariance}
\end{align}
In contrast, $\mathcal H_{p,m}^{-1}$ describes only the local curvature of the penalized objective. 
Equation~(\ref{eq:pen_covariance}) reduces to the usual inverse observed information when $\gamma=0$, 
but it neither corrects finite-sample shrinkage bias nor accounts for model and tuning selection.
Accordingly, the significance symbols in Table~3 are exploratory,
model-conditional summaries rather than selection-adjusted tests. 
For regression coefficients, we use the larger of the standard errors obtained from $\Psi$ and
Eq.~(\ref{eq:pen_covariance}), as described after Eq.~(18); this conservative rule is not a formal post-selection correction. 
Appendix~C gives the corresponding first-order asymptotic argument.

\textbf{Averaged pseudo-$R^2$}: 
Because the usual variance decomposition need not hold for spatial autoregressive models \citep[see, e.g.,][]{lesage2009,elhorst2014}, 
we use
$$
\bar R^2_{\mathrm{pseudo}}=K^{-1}\sum_{k=1}^K \mathrm{corr}(\by_k, \ \hat{\by}_k)^2.
$$ 
Pseudo-$R^2$ measures predictive performance only, 
whereas pAIC and pBIC additionally account for spatial dependence structures and model complexity.

The model-implied conditional covariance structure can also be used to construct principal components of the $K$ response variables. 
Appendix~D derives linear combinations that maximize the total variance aggregated over the spatial units, 
thereby providing a concise summary of their joint variation.

\section{Analysis of Synthetic and Empirical Data}  
We validated the proposed framework using synthetic and empirical data by fitting the eleven models shown in Figure~1.
The two sufficient identifiability conditions in Propositions~1 and~2 in Section 3.2 were also examined numerically for both datasets.
For Proposition~1, we computed the smallest singular value of $\big[ G_{\mu_0}, X \big]$ 
for the synthetic data and that of $\big[ G_{\widehat{\mu}}, X \big]$ for the empirical data, 
where $\widehat{\mu}$ denotes the fitted mean vector.
For Proposition~2, we computed the smallest singular value of the matrix 
$\big[ \operatorname{vec}(I_n), \operatorname{vec}(W), \operatorname{vec}(W'), \operatorname{vec}(W'W) \, \big]$.
All the resulting smallest singular values were positive, providing numerical support for the required full-rank 
and linear-independence conditions in the analyses considered here. 

All simulation datasets, true parameter values, empirical data sources, R scripts, 
and complete numerical results are publicly available in the project repository \cite{ohta2026a}.
Detailed instructions for reproducing all analyses are provided in the accompanying README documentation.

 \begin{table}[htbp]
\centering
\caption{True parameter values used in the simulation study}
\label{tab:true_parameter_values}
\renewcommand{\arraystretch}{1.35}
\begin{tabular}{c@{\ \ }c@{\ \ }c@{\ \ }c@{\ \ }c}
\toprule
Parameter block
  & Parameter
  & Weak
  & Moderate
  & Strong \\
\midrule
Spatial lag dependence
  & $R$ (three levels)
  & $\begin{pmatrix}
      0.12 & 0.04 \\
      0.03 & 0.10
    \end{pmatrix}$
  & $\begin{pmatrix}
      0.30 & 0.10 \\
      0.08 & 0.25
    \end{pmatrix}$
  & $\begin{pmatrix}
      0.55 & 0.15 \\
      0.12 & 0.45
    \end{pmatrix}$ \\
\addlinespace
Spatial error dependence
  & $\Lambda$ (three levels)
  & $\begin{pmatrix}
      0.10 & 0.03 \\
      0.02 & 0.08
    \end{pmatrix}$
  & $\begin{pmatrix}
      0.25 & 0.08 \\
      0.05 & 0.20
    \end{pmatrix}$
  & $\begin{pmatrix}
      0.50 & 0.12 \\
      0.10 & 0.40
    \end{pmatrix}$ \\
\addlinespace
\midrule
Temporal dependence
  & $A$ (fixed)
  & \multicolumn{3}{c}{
      $\begin{pmatrix}
        0.30 & 0.03 \\
        0.05 & 0.25
      \end{pmatrix}$} \\
\addlinespace
Error covariance
  & $\Sigma$ (fixed)
  & \multicolumn{3}{c}{
      $\begin{pmatrix}
        0.10 & 0.03 \\
        0.03 & 0.10
      \end{pmatrix}$} \\
\addlinespace
\multirow{2}{*}{Regression coefficients}
  & $\bbeta_1^*$ (fixed)
  & \multicolumn{3}{c}{$(\ \; 0.2,\ 1.2,\ -0.6,\ 0.8)'$} \\
  & $\bbeta_2^*$ (fixed)
  & \multicolumn{3}{c}{$(-0.1,\ 0.8,\ -0.3,\ 0.5)'$} \\
\bottomrule
\end{tabular}

\vspace{1mm}
\begin{minipage}{0.96\textwidth}
\footnotesize
\textit{Note:}
The regression coefficient vectors correspond to the intercept, two common
covariates, and one response-specific covariate, in that order.
For DGP 1111, both $R$ and $\Lambda$ take the values shown above.
For DGP 1011, $R$ takes the displayed values and $\Lambda=\mathbf{0}$.
For DGP 0111, $R=\mathbf{0}$ and $\Lambda$ takes the displayed values.
The values of $\bbeta_1^*$, $\bbeta_2^*$, $A$, and $\Sigma$ are 
common to all three spatial-dependence settings and all three DGPs.
\end{minipage}
\end{table}

\subsection{Validation with synthetic data}       
\label{sec:simulation}
We conducted a Monte Carlo experiment to evaluate parameter recovery and 
model-selection performance under different sample sizes and degrees of spatial dependence. 
Spatio-temporal bivariate data were generated on square lattices of sizes $10\times10$, $20\times20$, 
and $30\times30$, corresponding to $n=100$, 400, and 900 spatial units, respectively. 
For each lattice, the spatial weight matrix $W$ was constructed using queen contiguity and row-standardized.

Three data-generating processes (DGPs) were considered: 
the full MGNST model (1111), 
the spatial-lag model without spatial error dependence (1011), and 
the spatial-error model without spatial lag dependence (0111). 
For each DGP, the nonzero spatial parameter matrices were specified at weak, moderate, and strong levels. 
The experiment therefore comprised 27 combinations of DGP, sample size, and spatial-dependence strength. 
For each combination, 300 independent datasets were generated, giving 8,100 Monte Carlo replications in total.
The identifiability conditions were verified numerically for all simulated datasets. 
For example, for the full model (1111) at $n=100$, the minima of the smallest singular values of 
$[\,G_{\mu_0},X\,]$ over the 300 replications were 0.207, 0.155, and 0.048 
under weak, moderate, and strong spatial dependence, respectively.

The two response equations contained two common covariates and one response-specific covariate. 
Specifically, the design matrices were \[
 X_1^* = [\boldsymbol{1},\bx_1,\bx_2,\bx_3],
 \qquad
 X_2^* = [\boldsymbol{1},\bx_1,\bx_2,\bx_4],
 \qquad \1 = (1, ... , 1)': n \times 1,
\]
where $\bx_1$ and $\bx_2$ were common to both responses, whereas $\bx_3$ and $\bx_4$ were response-specific. 
Table~1 reports the true parameter values. 
The spatial matrices $R$ and $\Lambda$ varied across the three dependence levels, 
whereas $A$, $\Sigma$, $\bbeta_1^*$, and $\bbeta_2^*$ were held fixed.

\begin{figure}[p]
  \centering
  \begin{minipage}[c]{0.495\textwidth}
    \centering
    \includegraphics[
      width=\linewidth,
      height=0.25\textheight,
      keepaspectratio
    ]{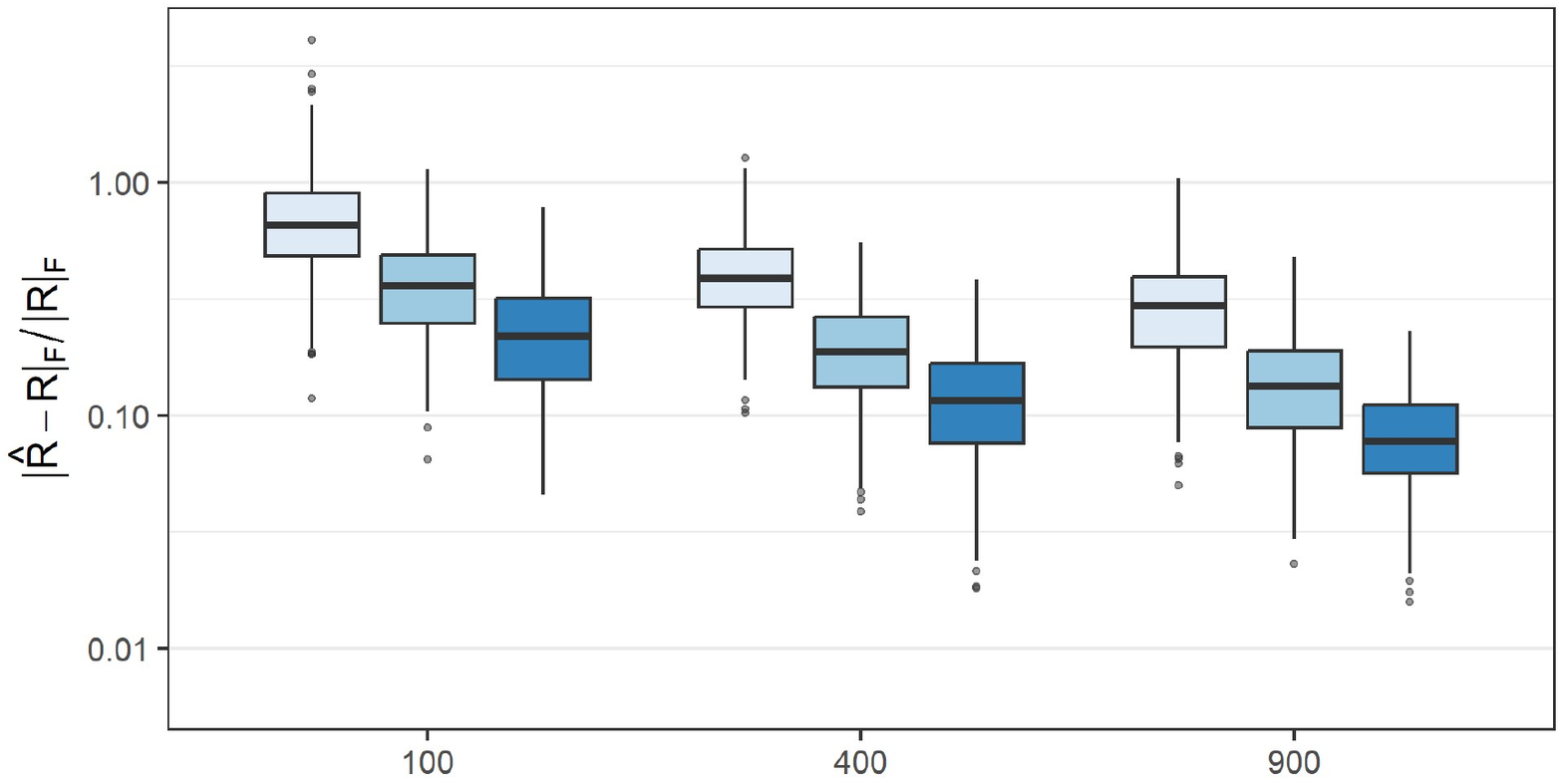}
  \\[-5mm] \hfil $R$ (Spatial lag matrix)
  \end{minipage}
  \hfill
  \begin{minipage}[c]{0.495\textwidth}
    \centering
    \includegraphics[
      width=\linewidth,
      height=0.25\textheight,
      keepaspectratio
    ]{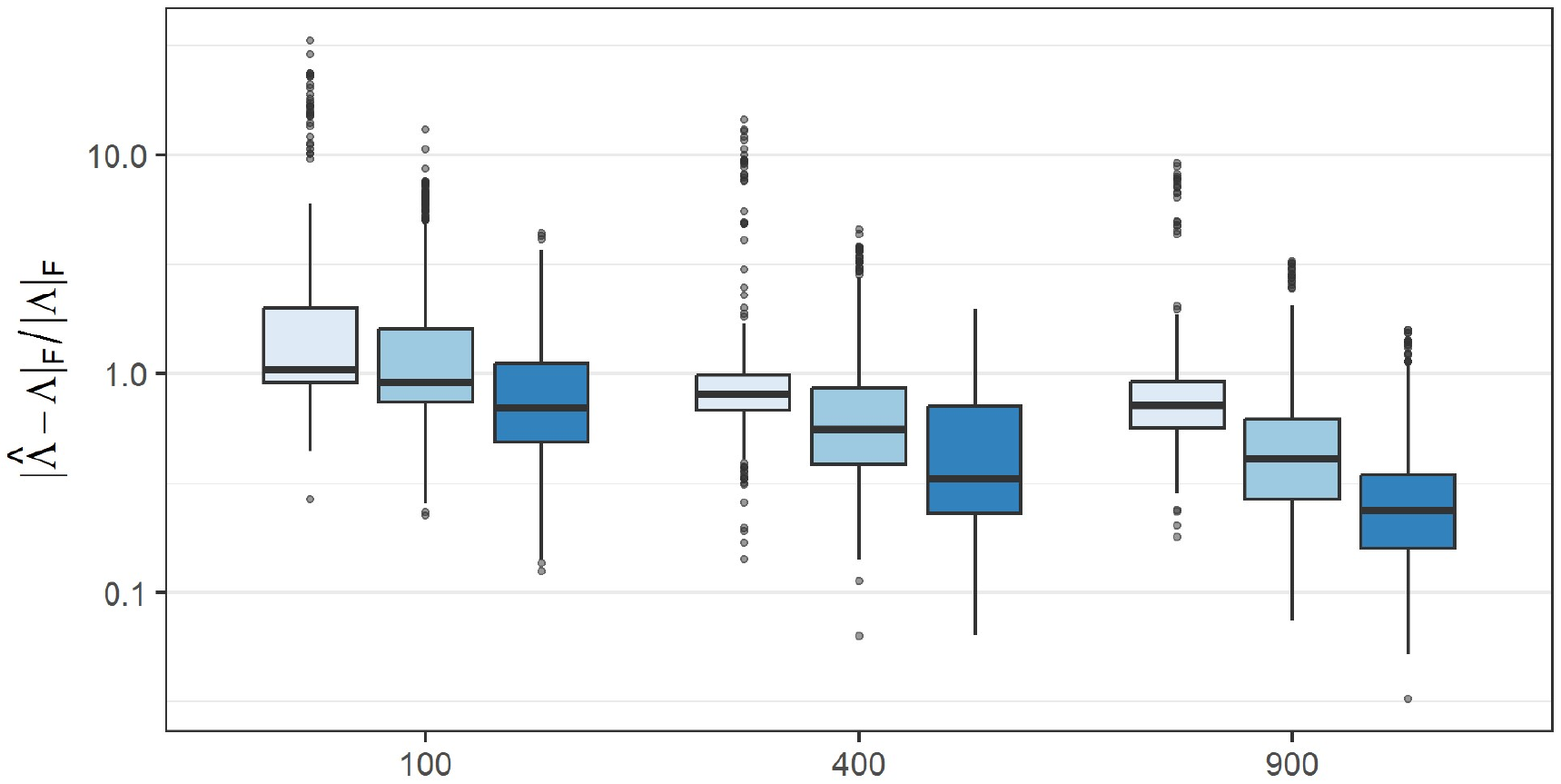}
  \\[-5mm] \hfil $\Lambda$ (Spatial error matrix)
  \end{minipage}


  \vspace{2mm}      
  \begin{minipage}[c]{0.495\textwidth}
    \centering
    \includegraphics[
      width=\linewidth,
      height=0.25\textheight,
      keepaspectratio
    ]{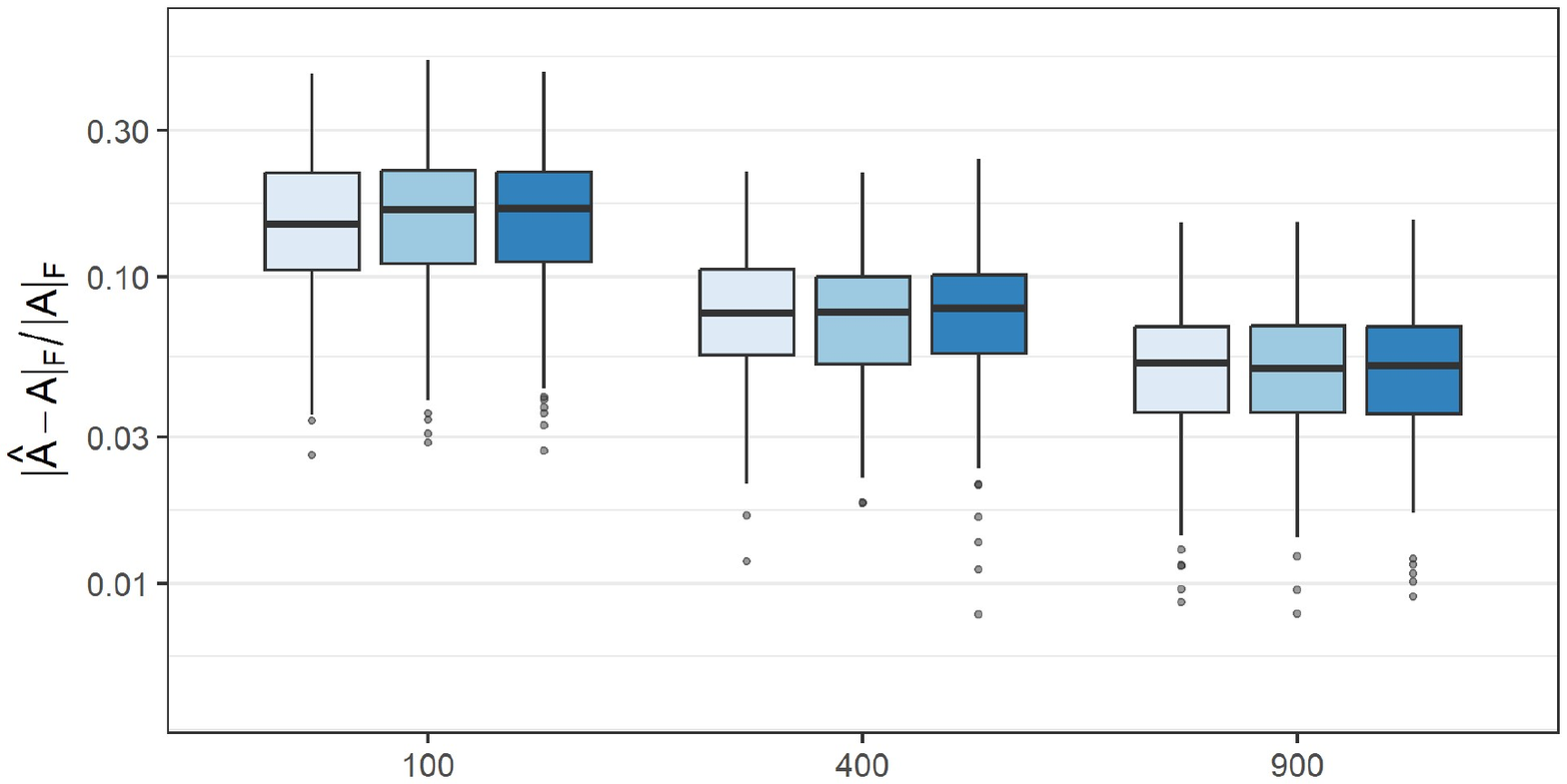}
  \\[-5mm] \hfil $A$ (AR(1) coefficients)
  \end{minipage}
  \hfil
  \begin{minipage}[c]{0.495\textwidth}
    \centering
    \includegraphics[
      width=\linewidth,
      height=0.25\textheight,
      keepaspectratio
    ]{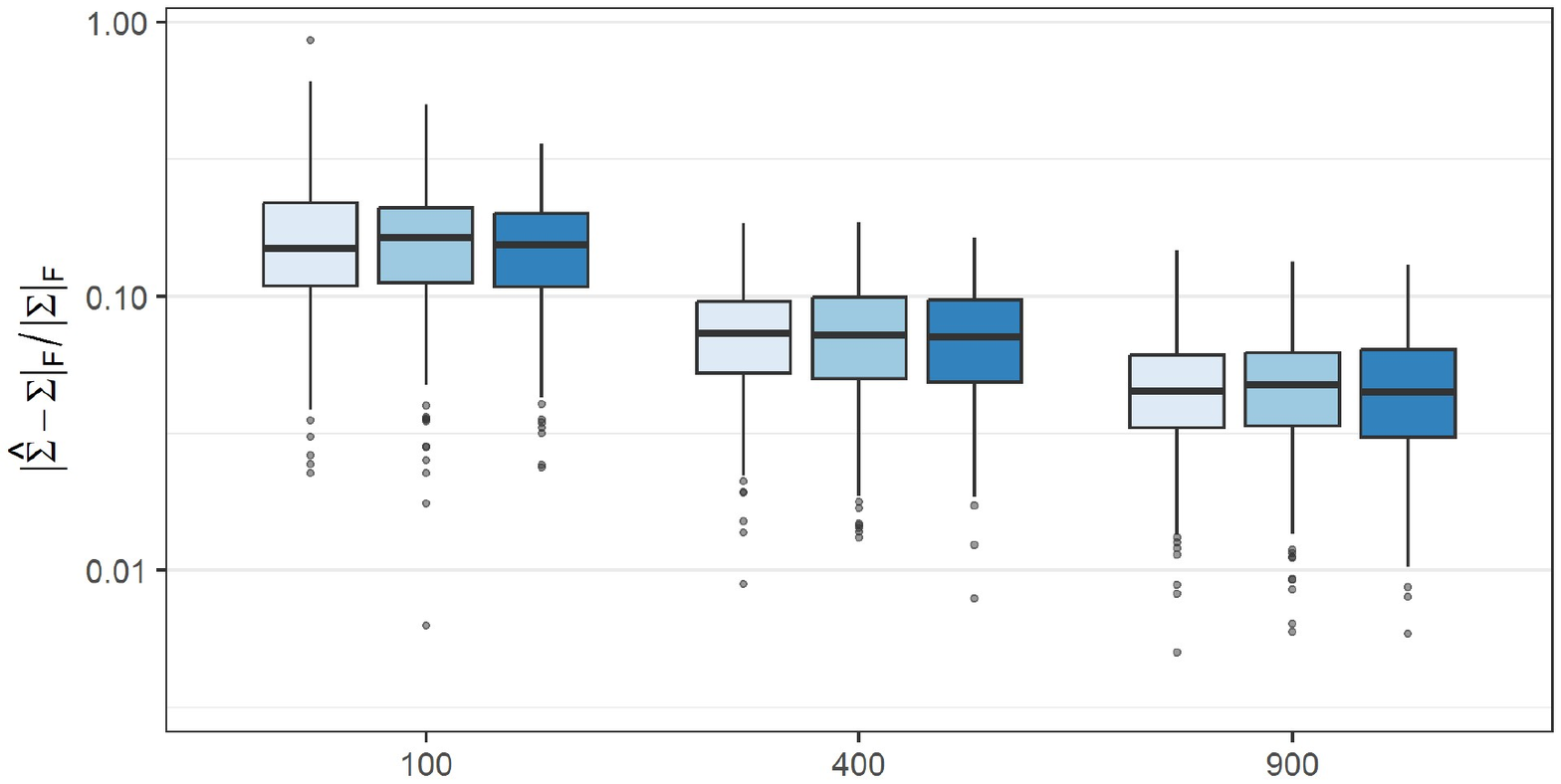}
  \\[-5mm] \hfil $\Sigma$ (Error covariance)
  \end{minipage}

  \vspace{2mm}      
  \begin{minipage}[c]{0.495\textwidth}
    \centering
    \includegraphics[
      width=\linewidth,
      height=0.25\textheight,
      keepaspectratio
    ]{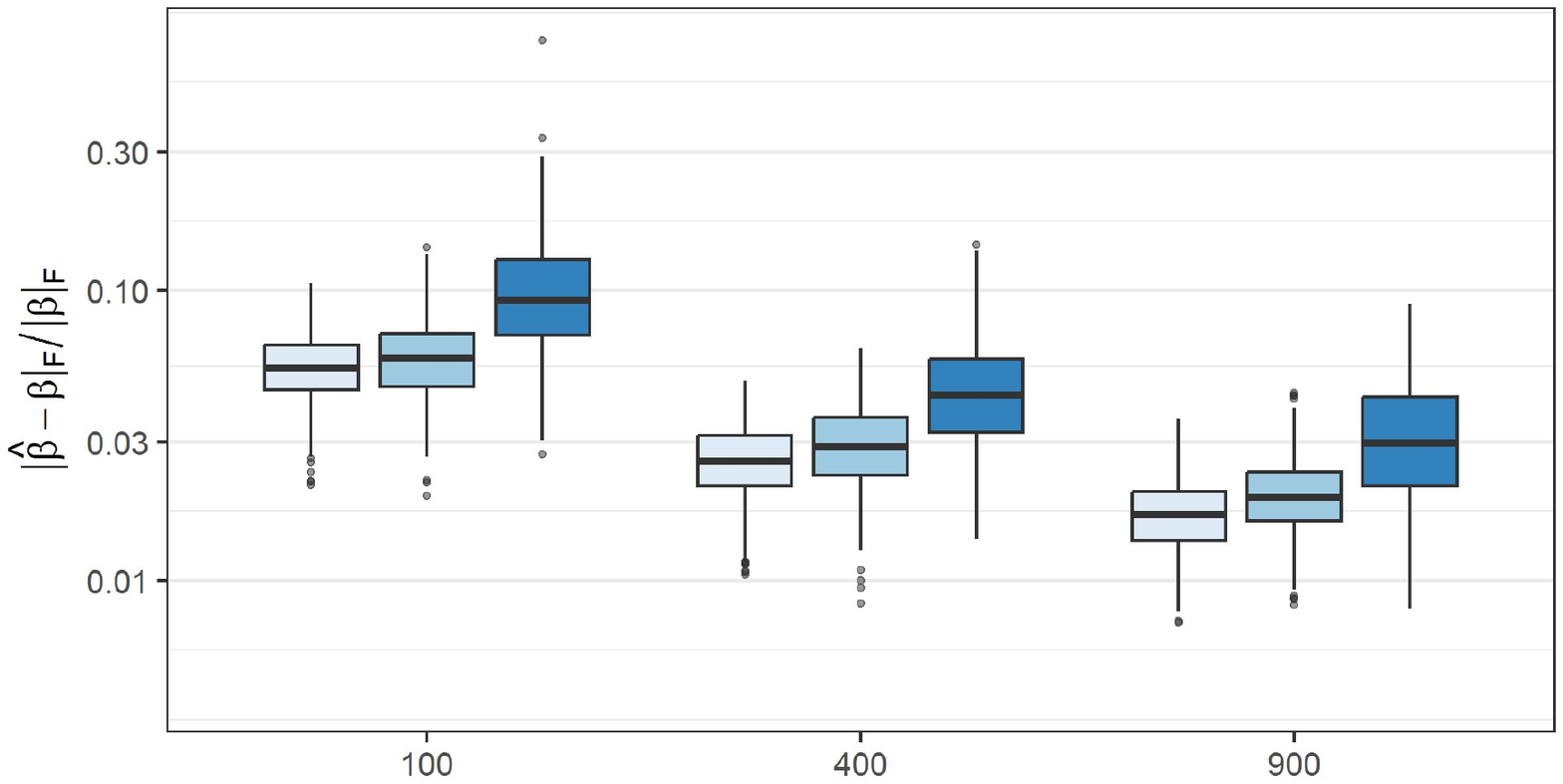}
  \\[-5mm] \hfil $\bbeta^*$ (Regression coefficients excluding AR(1))
  \end{minipage}
  \hfill
  \begin{minipage}[c]{0.495\textwidth}
    \centering
    \includegraphics[
      width=\linewidth,
      height=0.25\textheight,
      keepaspectratio
    ]{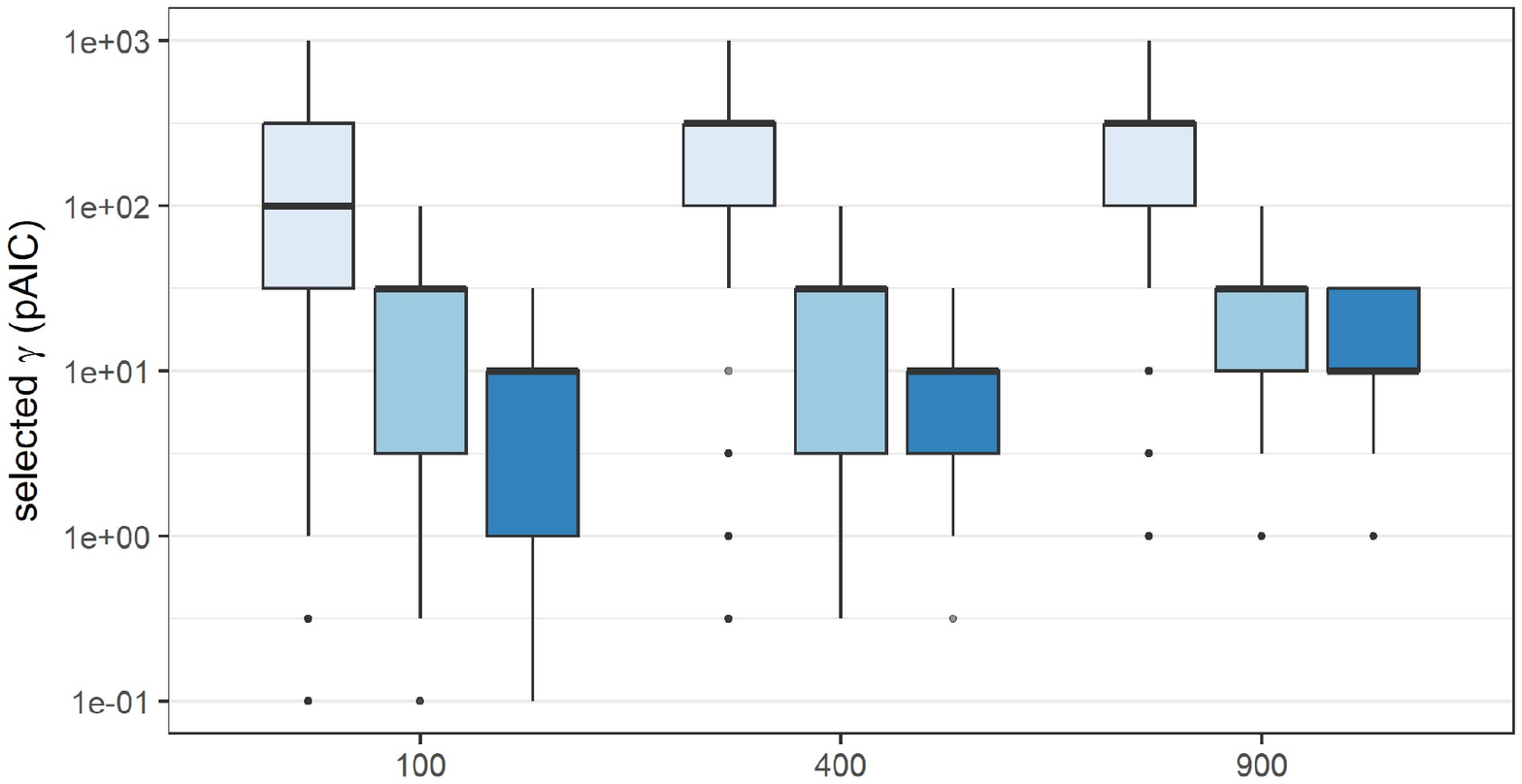}
  \\[-5mm] \hfil $\gamma$ frequencies selected by pAIC
  \end{minipage}

\caption{Accuracy of parameter estimation and distribution of the penalty weight based on 300 independent datasets under DGP (1111).
The first five panels show the relative Frobenius errors of the parameter estimates. 
From top left to bottom right, the panels correspond to 
$R$ (spatial lag matrix), 
$\Lambda$ (spatial error matrix),
$A$ (AR(1) coefficient matrix), $\Sigma$ (error covariance matrix),
$\bbeta^*$ (regression coefficients excluding the AR(1) coefficients),
and the empirical frequencies of the penalty weight 
$\gamma \in \{0, 10^{-2}, 10^{-1.5}, 10^{-1}, 10^{-0.5}, 1, 10^{0.5}, 10, 10^{1.5}, 100, 10^{2.5}, 1000 \}$ selected by pAIC. 
Within each sample size, the three shades represent weak, moderate, and strong spatial dependence.}
\label{fig:parameter_accuracy}
\end{figure}

 \begin{table}[htbp]
\centering
\caption{Model selection frequencies based on pAIC over 300 replications}
\label{tab:paic}
\begin{tabular}{c@{\ \ }c@{\ \ }c@{\ \ }c@{\ \ }c@{\ \ }c@{\ \ }c@{\ \ }c@{\ \ }c@{\ \ }c}
\toprule
\multirow{2}{*}{True model}
  & \multirow{2}{*}{Sample size}
  & \multirow{2}{*}{Spatial dependence}
  & \multicolumn{7}{c}{Selection frequency (\%)} \\
\cmidrule(lr){4-10}
  & & & 1111 & 0111 & 1011 & 0011 & d0dd & 0ddd & dddd \\
\midrule
\multirow{9}{*}{1111}
  & \multirow{3}{*}{100}
  & Weak       & \textbf{44.7}  & 8.0 & 43.0 & 0.7 & 2.7 & 0.3 & 0.7 \\
  & & Moderate & \textbf{60.7}  &  0  & 37.3 &  0  & 2.0 &  0  &  0  \\
  & & Strong   & \textbf{95.7}  &  0  & 4.3  &  0  &  0  &  0  &  0  \\
\cmidrule(lr){2-10}
  & \multirow{3}{*}{400}
  & Weak       & \textbf{46.7}  &  0  & 53.3 &  0  &  0  &  0  &  0  \\
  & & Moderate & \textbf{93.7}  &  0  & 6.3  &  0  &  0  &  0  &  0  \\
  & & Strong   & \textbf{100}   &  0  &  0   &  0  &  0  &  0  &  0  \\
\cmidrule(lr){2-10}
  & \multirow{3}{*}{900}
  & Weak       & \textbf{66.3}  &  0  & 33.7 &  0  &  0  &  0  &  0  \\
  & & Moderate & \textbf{100}   &  0  &  0   &  0  &  0  &  0  &  0  \\
  & & Strong   & \textbf{100}   &  0  &  0   &  0  &  0  &  0  &  0  \\
\midrule
\multirow{9}{*}{0111}
  & \multirow{3}{*}{100}
  & Weak       & 14.3 & \textbf{35.0} & 15.3 & 33.3 & 1.0 & 1.0 &  0  \\
  & & Moderate & 21.7 & \textbf{50.0} & 14.7 & 11.7 & 0.3 & 1.3 & 0.3 \\
  & & Strong   & 19.0 & \textbf{74.7} & 2.3  & 0.7  &  0  & 3.0 & 0.3 \\
\cmidrule(lr){2-10}
  & \multirow{3}{*}{400}
  & Weak       & 11.7 & \textbf{46.7} & 12.3 & 29.3 &  0  &  0  &  0  \\
  & & Moderate & 17.0 & \textbf{81.0} & 1.0  & 1.0  &  0  &  0  &  0  \\
  & & Strong   & 12.3 & \textbf{87.7} &  0   &  0   &  0  &  0  &  0  \\
\cmidrule(lr){2-10}
  & \multirow{3}{*}{900}
  & Weak       & 11.7 & \textbf{66.3} & 8.7 & 13.3 &  0  &  0  &  0  \\
  & & Moderate & 10.7 & \textbf{89.3} &  0  &  0   &  0  &  0  &  0  \\
  & & Strong   & 12.0 & \textbf{88.0} &  0  &  0   &  0  &  0  &  0  \\
\midrule
\multirow{9}{*}{1011}
  & \multirow{3}{*}{100}
  & Weak       & 36.3 & 6.7 & \textbf{50.7} & 2.0 & 4.0 &  0  & 0.3 \\
  & & Moderate & 46.0 &  0  & \textbf{51.3} &  0  & 2.0 &  0  & 0.7 \\
  & & Strong   & 49.0 &  0  & \textbf{50.3} &  0  & 0.7 &  0  &  0  \\
\cmidrule(lr){2-10}
  & \multirow{3}{*}{400}
  & Weak       & 39.7 &  0  & \textbf{60.3} &  0  &  0  &  0  &  0  \\
  & & Moderate & 30.3 &  0  & \textbf{69.7} &  0  &  0  &  0  &  0  \\
  & & Strong   & 28.0 &  0  & \textbf{72.0} &  0  &  0  &  0  &  0  \\
\cmidrule(lr){2-10}
  & \multirow{3}{*}{900}
  & Weak       & 31.3 &  0  & \textbf{68.7} &  0  &  0  &  0  &  0  \\
  & & Moderate & 31.3 &  0  & \textbf{68.7} &  0  &  0  &  0  &  0  \\
  & & Strong   & 23.0 &  0  & \textbf{77.0} &  0  &  0  &  0  &  0  \\
\bottomrule
\end{tabular}
\end{table}

For every generated dataset, the 11 candidate specifications summarized in Figure~1 were fitted 
using the estimation procedure described in Section~4.2. 
The spatial penalty weight $\gamma$ was selected from the candidate grid used in the numerical analysis, 
and the fitted models were compared using pAIC. 
Table~2 reports the percentage of the 300 replications in which each candidate model attained the smallest pAIC 
under each experimental condition. 
Thus, an entry in the column corresponding to the true model gives its correct-selection rate. 
Candidate models that were never selected as the pAIC minimizer are omitted from the table.

For DGPs (1111) and (0111), the correct-selection rates generally increased 
with sample size and the strength of the relevant spatial dependence.
For DGP  (1111), the rate increased from 44.7\% under weak dependence at $n=100$ 
to 100\% under moderate or strong dependence at $n=900$.
For DGP  (0111), the corresponding rates ranged from 35.0\%--74.7\% at $n=100$ to 66.3\%--89.3\% at $n=900$.
Under weak dependence, reduced models that omitted one or both spatial components remained frequent competitors, 
but their selection frequencies generally declined as the sample size increased.

Distinguishing DGP (1011) from the full model (1111) was more difficult.
The true-model selection rate was approximately 50\% at $n=100$ and increased to 68.7\%--77.0\% at $n=900$, 
with model (1111) accounting for most of the remaining selections. 
Across all three DGPs, diagonal and response-wise independent specifications were rarely selected 
and essentially disappeared at the larger sample sizes.

Figure~2 summarizes parameter recovery under DGP (1111) using estimates from the pAIC-selected models. 
For a parameter $B$, accuracy is measured by the relative Frobenius error \\ $\|\hat B-B\|_F \, \big/ \, \|B\|_F$.

The relative errors of the two spatial parameter matrices, $R$ and $\Lambda$, decreased 
as the sample size increased and also as the underlying spatial dependence became stronger.
This pattern is particularly pronounced for $\Lambda$, 
whose estimation was relatively unstable under weak dependence but improved substantially 
when the spatial-error signal became stronger.
Hence, both a larger number of spatial units and a stronger spatial signal facilitated identification 
and estimation of the spatial dependence parameters.

In contrast, the relative errors of $A$ and $\Sigma$ decreased mainly with sample size and showed little systematic variation 
across the three levels of spatial dependence.
This is consistent with the fact that their true values were held fixed across the spatial-dependence settings 
and suggests that their estimation accuracy is driven primarily by the amount of data 
rather than by the magnitudes of $R$ and $\Lambda$.

The regression coefficient vector $\bbeta^*$ in Figure~2 exhibited a different pattern.
Its relative error decreased with sample size but, within each sample size, tended to increase 
as the spatial dependence in $R$ and $\Lambda$ became stronger.
Thus, stronger spatial dependence improved recovery of the spatial parameter matrices themselves 
but was accompanied by less efficient estimation of the regression coefficients.
This pattern may reflect increasing difficulty in separating covariate effects from spatial effects 
as spatial propagation becomes stronger, 
although its precise mechanism requires further investigation.

The lower-right panel of Figure~2 reports the empirical frequencies of the penalty weights $\gamma$ selected by pAIC.
The boxplots display the nine values of $\gamma$ that were selected at least once from the twelve candidate values.
Their selection frequencies varied systematically across sample sizes and spatial-dependence levels, 
indicating that the degree of penalization adapted to the amount of information 
and the strength of the spatial signal rather than remaining fixed across the simulation settings.
Overall, model selection was most difficult under weak spatial dependence and
when comparing nested models that differed in only one spatial component.

\subsection{Empirical analysis for the Kansai region}   

The proposed MGNST framework was applied to municipal-level socioeconomic data for 198 municipalities 
in the Kansai region of Japan observed in 2015 and 2020. 
The response variables were 
$\by_1$ = the standardized proportion of the population aged 65 years or older ({\bf older-population share}), and
$\by_2$ = the standardized proportion of workers employed in the secondary sector ({\bf secondary-sector employment}). 
The common explanatory variables were
$\bx_1$ = the log of taxable income per capita ({\bf income}) and
$\bx_2$ = $\log(\text{number of foreign residents}+1)$ ({\bf foreign residents}).
The response-specific covariates were
$\bx_3$ = the net migration rate ({\bf net migration}) for the first response
and
$\bx_4$ = the commuter inflow rate ({\bf commuter inflow}) for the second response.
A row-standardized municipal adjacency matrix was used as the spatial weight matrix. 
Figure~3 illustrates the spatial distributions of the two response variables in 2020. 
Both variables exhibit strong positive spatial autocorrelation, with Moran's $I$ values of 0.629 and 0.689, respectively; 
the corresponding $p$-values were calculated using a randomization-based normal approximation.

\begin{figure}[H]                                
  \centering
  \includegraphics[
    width=0.78\linewidth,
    keepaspectratio,
    pagebox=mediabox,
    trim = 0 65 0 40, clip    
  ]{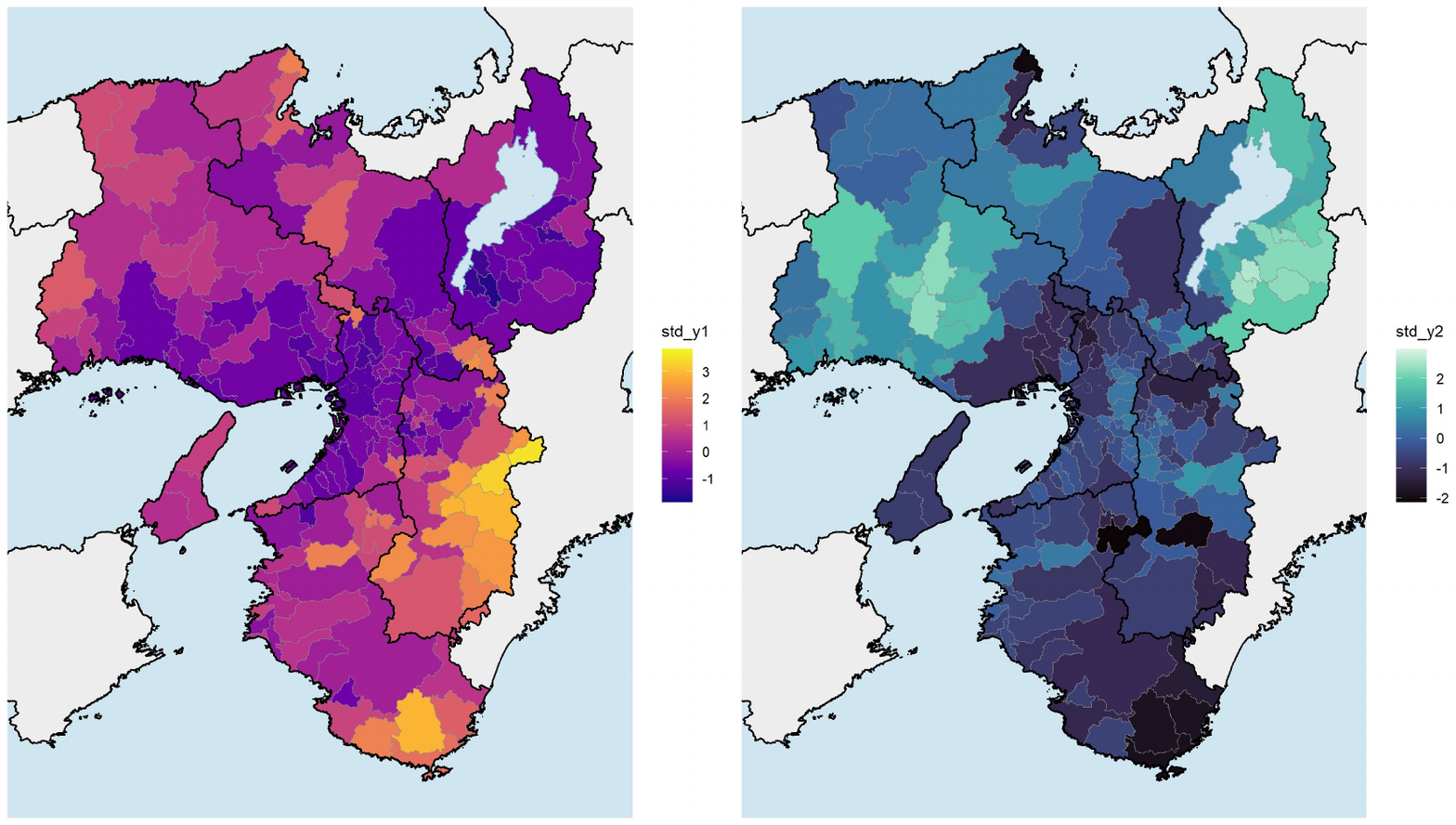}
  \caption{Choropleth maps of the responses in 2020 for 198 municipalities in the Kansai region of Japan.
    Left panel: $\by_1=\mbox{older-population share}$,
    $I=0.629$, $p\mbox{-value}<10^{-41}$;
    right panel: $\by_2=\text{secondary-sector employment}$,
    $I=0.689$, $p\mbox{-value}<10^{-49}$.}
  \label{fig:empirical_observed}
\end{figure}

The identifiability conditions were also examined numerically for the empirical data. 
The unknown true mean vector $\bmu_0$ was replaced by the fitted mean vector $\hat{\bmu}$ obtained from the pAIC-selected full model (1111), 
with $\gamma$ also selected by pAIC. 
The minimum singular value of $[\,G_{\hat{\mu}},X\,]$ was 0.00777. 
Furthermore, the minimum singular value of 
$[\operatorname{vec}(I_n),\operatorname{vec}(W),\operatorname{vec}(W'),\operatorname{vec}(W' W)]$ was 0.417. 
These positive values provide numerical support for the full-rank and linear-independence conditions 
in Propositions~1 and~2 for the empirical data.

\begin{sidewaystable}[htbp] 
\centering
\caption{Estimation results and model comparison based on pAIC for the empirical data with $K = 2$ and $n = 198$\\
Covariates observed in 2020: 
$\bx_1$ = log(Taxable income per capita),
$\bx_2$ = log(Foreign resident population + 1), 
$\bx_3$ = Net migration rate, 
$\bx_4$ = Commuter inflow rate.\\
}
\label{tab:empirical_aic}
\small

\begin{tabular}{@{\ \ \ }c  @{\ }rc@{\ } rc@{\ } rc@{\ } rc@{\ } rc@{\ } rc@{\ }  rc@{\ } rc@{\ } rc@{\ } rc@{\ } rc@{\ } rc@{\ }   rc@{\ \ \ }}
\toprule
Param.             & \multic{1111} & \multic{0111} & \multic{1011} & \multic{1101} & \multic{0101}  & \multic{1001} & \multic{d0dd} & \multic{0ddd} & \multic{dddd} & \multic{0011} & \multic{000d} \\
\midrule
\multicolumn{23}{c}{\textbf{Regression coefficients for $\by_1$ : Proportion of population aged 65 years or older in 2020}}\\
1                  & $\phm2.605$  & $*$     & $\phm1.982$  & $\cdot$ & $\phm1.806$    & $\cdot$ & $\phm25.419$ & \sss    & $\phm28.274$ & \sss    & $\phm18.930$ & \sss    & $\phm1.762$  & $\cdot$ & $\phm1.949$  & $\cdot$ & $\phm2.373$  & $*$     & $\phm1.799$  & $\cdot$ & $\phm31.144$ & \sss    \\
$\bx_1$            & $-0.276$     & $\cdot$ & $-0.199$     & $\piii$ & $-0.185$       & $\piii$ & $-2.970$     & \sss    & $-3.331$     & \sss    & $-2.191$     & \sss    & $-0.178$     & $\piii$ & $-0.194$     & $\piii$ & $-0.242$     & $\cdot$ & $-0.184$     & $\piii$ & $-3.569$     & \sss    \\
$\bx_2$            & $-0.013$     & $\piii$ & $-0.013$     & $\piii$ & $-0.005$       & $\piii$ & $-0.236$     & $**$    & $-0.220$     & $**$    & $-0.205$     & $**$    & $-0.006$     & $\piii$ & $-0.014$     & $\piii$ & $-0.018$     & $\piii$ & $-0.005$     & $\piii$ & $-0.375$     & \sss    \\
$\bx_3$            & $-0.150$     & \sss    & $-0.150$     & \sss    & $-0.157$       & \sss    & $-0.390$     & \sss    & $-0.367$     & \sss    & $-0.383$     & \sss    & $-0.158$     & \sss    & $-0.148$     & \sss    & $-0.151$     & \sss    & $-0.157$     & \sss    & $-0.539$     & \sss    \\[2mm]
\multicolumn{23}{c}{\textbf{Regression coefficients for $\by_2$: Proportion of secondary sector employment in 2020}}\\
1                  & $\phm3.653$  & \sss    & $\phm3.156$  & $**$    & $\phm3.808$    & \sss    & $\phm12.389$ & \sss    & $\phm13.589$ & \sss    & $\phm12.541$ & \sss    & $\phm1.471$  & $*$     & $\phm1.616$  & $*$     & $\phm1.617$  & $*$     & $\phm3.568$  & \sss    & $\phm9.738$  & $**$    \\
$\bx_1$            & $-0.491$     & \sss    & $-0.431$     & \sss    & $-0.506$       & \sss    & $-1.731$     & \sss    & $-1.860$     & \sss    & $-1.752$     & \sss    & $-0.224$     & $**$    & $-0.248$     & $*$     & $-0.246$     & $*$     & $-0.478$     & \sss    & $-1.711$     & \sss    \\
$\bx_2$            & $\phm0.020$  & $\piii$ & $\phm0.021$  & $\piii$ & $\phm0.014$    & $\piii$ & $\phm0.201$  & $*$     & $\phm0.165$  & $**$    & $\phm0.203$  & $**$    & $\phm0.026$  & $\cdot$ & $\phm0.032$  & $*$     & $\phm0.030$  & $*$     & $\phm0.015$  & $\piii$ & $\phm0.530$  & \sss    \\
$\bx_4$            & $\phm0.001$  & $\cdot$ & $\phm0.001$  & $*$     & $\phm0.001$    & $\cdot$ & $\phm0.004$  & $*$     & $\phm0.004$  & $*$     & $\phm0.004$  & $*$     & $\phm0.001$  & $*$     & $\phm0.001$  & $*$     & $\phm0.001$  & $*$     & $\phm0.001$  & $*$     & $\phm0.008$  & $**$    \\[2mm]
\multicolumn{23}{c}{\textbf{Autoregressive coefficients}}\\
$\alpha_{11}$      & $\phm1.037$  & \sss    & $\phm1.024$  & \sss    & $\phm1.017$    & \sss    &              & $\piii$ &              & $\piii$ &              & $\piii$ & $\phm1.017$  & \sss    & $\phm1.025$  & \sss    & $\phm1.039$  & \sss    & $\phm1.014$  & \sss    &              & $\piii$ \\
$\alpha_{12}$      & $-0.021$     & $\piii$ & $-0.004$     & $\piii$ & $-0.013$       & $\piii$ &              & $\piii$ &              & $\piii$ &              & $\piii$ &              & $\piii$ &              & $\piii$ &              & $\piii$ & $-0.002$     & $\piii$ &              & $\piii$ \\
$\alpha_{21}$      & $-0.028$     & $\piii$ & $-0.047$     & $*$     & $-0.042$       & $\cdot$ &              & $\piii$ &              & $\piii$ &              & $\piii$ &              & $\piii$ &              & $\piii$ &              & $\piii$ & $-0.058$     & \sss    &              & $\piii$ \\
$\alpha_{22}$      & $\phm0.917$  & \sss    & $\phm0.937$  & \sss    & $\phm0.924$    & \sss    &              & $\piii$ &              & $\piii$ &              & $\piii$ & $\phm0.919$  & \sss    & $\phm0.942$  & \sss    & $\phm0.928$  & \sss    & $\phm0.947$  & \sss    &              & $\piii$ \\[2mm]
\multicolumn{23}{c}{\textbf{Spatial coefficients}}\\
$\rho_{11}$        & $-0.058$     & $*$     &              & $\piii$ & $-0.002$       & $\piii$ & $\phm0.312$  & \sss    &              & $\piii$ & $\phm0.521$  & \sss    & $-0.003$     & $\piii$ &              & $\piii$ & $-0.055$     & $\cdot$ &              & $\piii$ &              & $\piii$ \\
$\rho_{12}$        & $\phm0.012$  & $\piii$ &              & $\piii$ & $\phm0.015$    & $\piii$ & $-0.103$     & $*$     &              & $\piii$ & $-0.035$     & $\piii$ & $\phm     $  & $\piii$ &              & $\piii$ & $\phm     $  & $\piii$ &              & $\piii$ &              & $\piii$ \\
$\rho_{21}$        & $-0.044$     & $\cdot$ &              & $\piii$ & $-0.026$       & \sss    & $-0.208$     & \sss    &              & $\piii$ & $-0.216$     & \sss    & $\phm     $  & $\piii$ &              & $\piii$ & $\phm     $  & $\piii$ &              & $\piii$ &              & $\piii$ \\
$\rho_{22}$        & $\phm0.033$  & $\piii$ &              & $\piii$ & $\phm0.032$    & $**$    & $\phm0.763$  & \sss    &              & $\piii$ & $\phm0.735$  & \sss    & $\phm0.061$  & $*$     &              & $\piii$ & $\phm0.036$  & $\piii$ &              & $\piii$ &              & $\piii$ \\
$\lambda_{11}$     & $\phm0.516$  & \sss    & $\phm0.479$  & \sss    &                & $\piii$ & $\phm0.334$  & \sss    & $\phm0.673$  & \sss    &              & $\piii$ &              & $\piii$ & $\phm0.490$  & \sss    & $\phm0.535$  & \sss    &              & $\piii$ &              & $\piii$ \\
$\lambda_{12}$     & $\phm0.111$  & $*$     & $-0.137$     & $*$     &                & $\piii$ & $\phm0.059$  & $\piii$ & $-0.002$     & $\piii$ &              & $\piii$ &              & $\piii$ & $\phm     $  & $\piii$ & $\phm     $  & $\piii$ &              & $\piii$ &              & $\piii$ \\
$\lambda_{21}$     & $-0.026$     & $\piii$ & $\phm0.151$  & $*$     &                & $\piii$ & $\phm0.087$  & $\piii$ & $-0.238$     & $*$     &              & $\piii$ &              & $\piii$ & $\phm     $  & $\piii$ & $\phm     $  & $\piii$ &              & $\piii$ &              & $\piii$ \\
$\lambda_{22}$     & $\phm0.303$  & $**$    & $\phm0.374$  & \sss    &                & $\piii$ & $-0.134$     & $\cdot$ & $\phm0.782$  & \sss    &              & $\piii$ &              & $\piii$ & $\phm0.443$  & \sss    & $\phm0.398$  & \sss    &              & $\piii$ &              & $\piii$ \\[2mm]
\multicolumn{23}{c}{\textbf{Error variance/covariance}}\\
$\sigma_{11}$      & $\phm0.022$  & $\piii$ & $\phm0.023$  & $\piii$ & $\phm0.027$    & $\piii$ & $\phm0.268$  & $\piii$ & $\phm0.263$  & $\piii$ & $\phm0.273$  & $\piii$ & $\phm0.027$  & $\piii$ & $\phm0.023$  & $\piii$ & $\phm0.022$  & $\piii$ & $\phm0.027$  & $\piii$ & $\phm0.420$  & $\piii$ \\
$\sigma_{12}$      & $-0.002$     & $\piii$ & $-0.001$     & $\piii$ & $-0.001$       & $\piii$ & $-0.024$     & $\piii$ & $-0.035$     & $\piii$ & $-0.024$     & $\piii$ & $\phm     $  & $\piii$ & $\phm     $  & $\piii$ & $\phm     $  & $\piii$ & $-0.001$     & $\piii$ & $\phm     $  & $\piii$ \\
$\sigma_{22}$      & $\phm0.020$  & $\piii$ & $\phm0.020$  & $\piii$ & $\phm0.021$    & $\piii$ & $\phm0.244$  & $\piii$ & $\phm0.259$  & $\piii$ & $\phm0.253$  & $\piii$ & $\phm0.022$  & $\piii$ & $\phm0.020$  & $\piii$ & $\phm0.020$  & $\piii$ & $\phm0.022$  & $\piii$ & $\phm0.742$  & $\piii$ \\[2mm]
\multicolumn{23}{c}{\textbf{Model assessment}}\\
AIC                & $-350.2$     & $\piii$ & $-349.0$     & $\piii$ & $-318.5$       & $\piii$ & $\phm667.4$  & $\piii$ & $\phm676.9$  & $\piii$ & $\phm667.2$  & $\piii$ & $-315.0$     & $\piii$ & $-352.9$     & $\piii$ & $-354.4$     & $\piii$ & $-318.2$     & $\piii$ & $\phm904.8$  & $\piii$ \\
d.f.               & $\phm23$     & $\piii$ & $\phm19$     & $\piii$ & $\phm19$       & $\piii$ & $\phm19$     & $\piii$ & $\phm15$     & $\piii$ & $\phm15$     & $\piii$ & $\phm14$     & $\piii$ & $\phm14$     & $\piii$ & $\phm16$     & $\piii$ & $\phm15$     & $\piii$ & $\phm10$     & $\piii$ \\
$\gamma_{\tiny\rm pAIC}$& $\phm7.079$  & $\piii$ & $\phm5.623$  & $\piii$ & $\phm1258.9$   & $\piii$ & $\phm17.783$ & $\piii$ & $\phm4.467$  & $\piii$ & $\phm6.310$  & $\piii$ & $\phm0$      & $\piii$ & $\phm0$      & $\piii$ & $\phm0$      & $\piii$ & $\phm0$      & $\piii$ & $\phm0$      & $\piii$ \\
$d_{\mathrm{eff}}$ & $\phm17.009$ & $\piii$ & $\phm15.426$ & $\piii$ & $\phm17.211$   & $\piii$ & $\phm16.518$ & $\piii$ & $\phm14.879$ & $\piii$ & $\phm14.911$ & $\piii$ & $\phm14$     & $\piii$ & $\phm14    $ & $\piii$ & $\phm16    $ & $\piii$ & $\phm15$     & $\piii$ & $\phm10$     & $\piii$ \\
pAIC               & $-362.2$     & $\piii$ & $-356.2$     & $\piii$ & $-322.0$       & $\piii$ & $\phm662.5$  & $\piii$ & $\phm676.6$  & $\piii$ & $\phm667.0$  & $\piii$ & $-315.0$     & $\piii$ & $-352.9$     & $\piii$ & $-354.4$     & $\piii$ & $-318.2$     & $\piii$ & $\phm904.8$  & $\piii$ \\
$\bar{R}^2_{\mbox{\tiny pseudo}}$& $\phm0.980$  & $\piii$ & $\phm0.979$  & $\piii$ & $\phm0.977$    & $\piii$ & $\phm0.755$  & $\piii$ & $\phm0.755$  & $\piii$ & $\phm0.745$  & $\piii$ & $\phm0.976$  & $\piii$ & $\phm0.979$  & $\piii$ & $\phm0.980$  & $\piii$ & $\phm0.976$  & $\piii$ & $\phm0.438$  & $\piii$ \\[1mm]
\bottomrule
\end{tabular}
\\[1mm]
The tuning parameter $\gamma$ was selected using a two-stage grid search. 
In the first stage, the value $\gamma^\ast$ that minimized pAIC was identified over the coarse grid 
$
\{0\}\cup
\{10^{-2},10^{-1.5},10^{-1},10^{-0.5},1,10^{0.5},10, 10^{1.5},10^2,\ldots,10^6\}.
$
In the second stage, a finer search was conducted between the two coarse-grid points adjacent to $\gamma^\ast$, using increments of $0.05$ on the $\log_{10}$ scale.
Significance codes: ${***}\ p<0.001$, ${**}\ p<0.01$, ${*}\ p<0.05$, ${.}\ p<0.10$.
\end{sidewaystable}

Table~3 summarizes the estimation results and model comparison based on the proposed pAIC.
The first column lists the explanatory variables, AR(1) regression coefficients,
spatial parameters, and model evaluation measures, 
while the remaining columns report the corresponding estimates and measures for the 11 candidate models.
For each regression coefficient, statistical significance is assessed conservatively 
using the larger of the two standard-error estimates obtained from 
$\Psi$ in Eq.~(18) and the covariance approximation in Eq.~(21).
For the spatial parameter estimates, including those in the response-wise independent model specifications, 
statistical significance is assessed using the standard errors obtained from Eq.~(21).
Some estimated AR(1) coefficients exceeded one.
This can occur because the 2020 response variables were regressed on their corresponding 2015 values 
using only a single temporal transition; 
therefore, these coefficients represent associations between the two observation years rather than estimates from a long time series, 
and values exceeding one do not by themselves indicate explosive temporal dynamics.

Among the candidate models, the full MGNST model (1111) attained the smallest pAIC value of $-362.2$, 
followed by the multivariate spatial error model (0111) with $-356.2$. 
Both models retained full temporal and cross-equation covariance structures, 
and the response-wise independent models had larger pAIC values. 
These results support temporal persistence and spatial error dependence, 
while the improvement of model (1111) over model (0111) suggests additional explanatory value from spatial lag interactions.
Moreover, all highly ranked models retained temporal autoregressive terms, supporting substantial temporal persistence 
in the two socioeconomic responses.

The estimated regression coefficients were generally stable across the highest-ranked models. 
Net migration was consistently negatively associated with older-population share, 
and income was also negatively associated with secondary-sector employment. 
These associations remained statistically significant after spatial, temporal, 
and cross-equation dependence had been taken into account, 
indicating that they were not artifacts of the particular dependence specification selected.

Figure~4 provides an additional diagnostic based on Moran's $I$ for the estimated innovations,
defined as $\hat{\bepsilon} = S_{\hat{\Lambda}}\bigl(S_{\hat{R}}\by-X\hat{\bbeta}\bigr)$.
After fitting the pAIC-selected MGNST model, neither innovation exhibited significant residual spatial autocorrelation. 
Compared with the strong positive spatial autocorrelation in the original responses (Figure~3),
this result indicates that the fitted model adequately captured the spatial dependence in both responses.

\begin{figure}[H]                                 
  \centering
  \includegraphics[
    width=0.78\linewidth,
    keepaspectratio,
    pagebox=mediabox,
    trim = 0 66 0 20, clip    
  ]{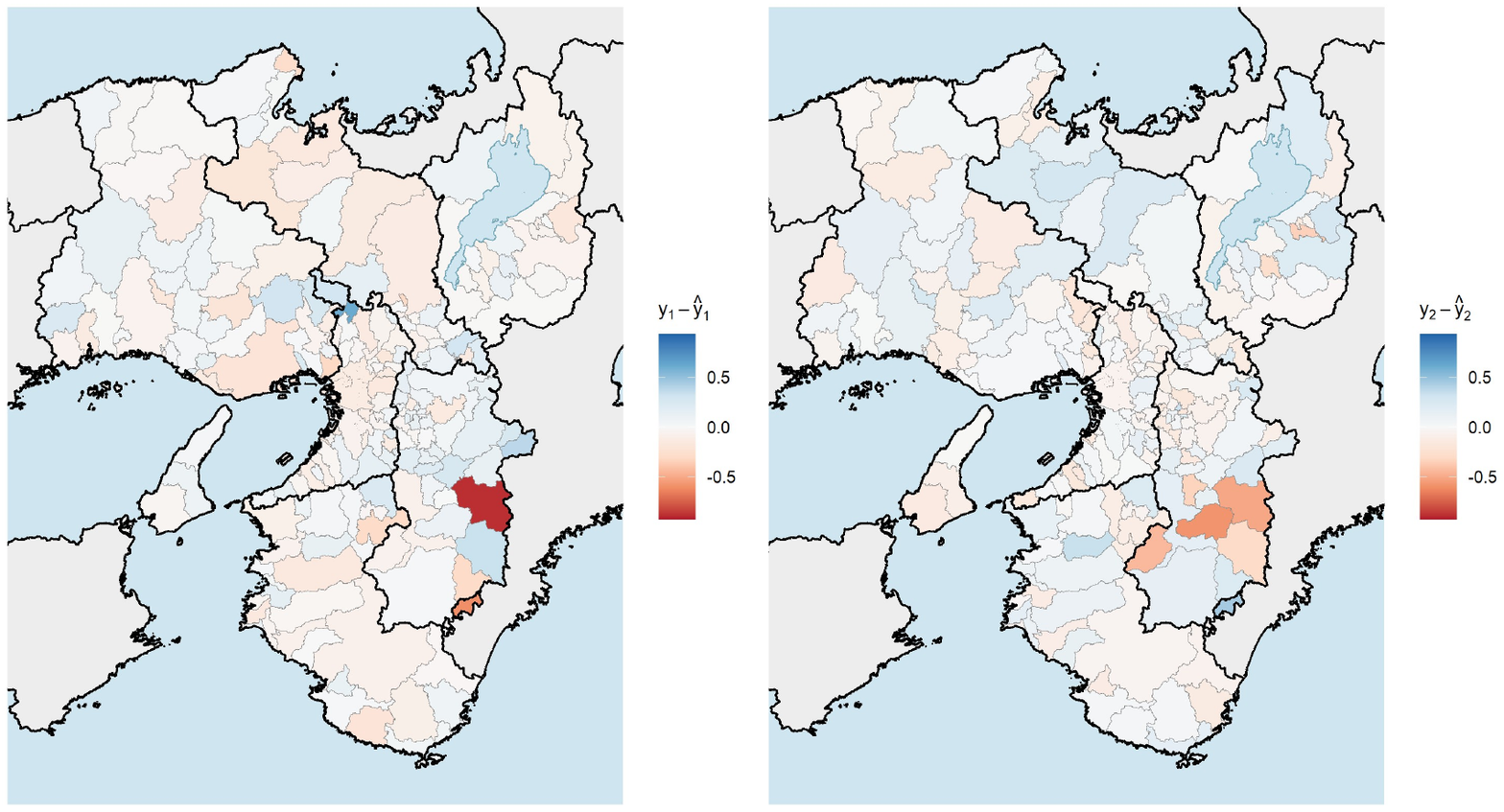}
  \caption{Choropleth maps of the estimated innovations in 2020 from the pAIC-selected MGNST model (1111).
    Left panel: older-population share, $\hat{\bepsilon}_1$, $I=-0.017$, $p=0.315$;
    right panel: secondary-sector employment, $\hat{\bepsilon}_2$, $I=0.022$, $p=0.278$.}
  \label{fig:empirical_residual}
\end{figure}

For comparison, pBIC, whose complete results are available in \cite{ohta2026a}, 
selected the simpler response-wise independent spatial error model (0ddd), 
followed by the multivariate spatial error model (0111) and the full MGNST model (1111). 
The difference between the pAIC and pBIC rankings reflects their different complexity penalties: 
pAIC favors the additional multivariate spatial structure of model (1111), 
whereas pBIC favors the more parsimonious model (0ddd). 
Nevertheless, both criteria retain temporal autoregressive and spatial error components, 
providing consistent evidence for temporal persistence and spatial error dependence in the data.

\section{Conclusion and Discussion}                         
This paper developed the statistical foundations of a multivariate general nesting spatio-temporal (MGNST) regression framework 
for analyzing processes that exhibit simultaneous spatial, temporal, and cross-equation dependence.
The framework jointly incorporates spatial lag effects, spatial error dependence, temporal autoregressive dynamics, 
and cross-equation covariance within a unified likelihood-based formulation.

From a theoretical perspective, sufficient identifiability conditions were established using \\ 
instrumental-variable rank conditions, 
and a penalized likelihood approach was introduced to stabilize estimation of spatial dependence parameters.
The resulting effective degrees of freedom provided a principled foundation for model comparison through penalized information criteria.

Simulation experiments showed that model selection and parameter recovery generally improved with sample size and spatial-signal strength. 
Closely nested spatial specifications remained more difficult to distinguish, particularly under weak dependence, 
whereas diagonal and response-wise independent models were rarely selected at larger sample sizes.

In the empirical analysis, pAIC selected the full MGNST model, 
whereas pBIC selected a response-wise independent spatial error model. 
Despite this difference in complexity, both criteria supported temporal persistence and spatial error dependence. 
The pAIC-selected model also reduced the strong spatial autocorrelation in the responses to negligible levels in the estimated innovations. 
Taken together, these results illustrate how the eleven specifications generated by restrictions on $(R,\Lambda,A,\Sigma)$ 
provide a common basis for comparing alternative multivariate dependence structures.

The methodology developed in this paper focuses on the single-transition conditional model $f(\by_t\mid\by_{t-1},X_t)$. 
With observations at multiple time points, the same specification yields the conditional likelihood
$\prod_{t=2}^T f(\by_t\mid\by_{t-1},X_t)$, 
while higher-order AR($s$) dependence can be accommodated by augmenting the design matrix with additional lagged responses.
Developing estimation and model-selection procedures that fully exploit these extensions remains an important topic for future research.

The proposed framework is applicable to any number $K$ of response variables.
Although the penalized likelihood approach stabilizes estimation, more flexible regularization 
may further improve statistical and computational performance.
A natural refinement is to introduce separate tuning parameters $\gamma_R$ and $\gamma_\Lambda$ 
for the quadratic penalties on $R$ and $\Lambda$ and develop data-driven methods for their joint selection. 
Further directions include a general-purpose software package, 
as well as asymptotic theory and structured or sparse regularization for large-$n$ and large-$K$ settings.

\begin{acks}[Acknowledgments]           
The first author gratefully acknowledges the continued administrative
support provided by Nagasaki University and Chuo University. 
This support enabled him to administer the JSPS KAKENHI project following his retirement.
This study was supported by JSPS KAKENHI Grant-in-Aid for Scientific Research (Project No.~24K14857).
\end{acks}



\section*{Appendix A: Proofs of the Identifiability Propositions}     
In this appendix, we prove the two propositions described in Section 3.2.

\smallskip 
{\bf Proof of Proposition 1}. \ \ 
A straightforward calculation gives
\begin{align}
 ( R_0 \otimes W ) \; \bmu_0 \ = \ G_{\mu_0} \operatorname{vec}(R_0).
\end{align}
Note that the true parameters satisfy $ E( S_{R_0} \, \by) = S_{R_0} \, \bmu_0 = X \, \bbeta_0$.
Suppose another parameter pair $(R,\bbeta)$ produces the same population mean vector.
It holds that $S_R \, \bmu_0 = X \bbeta$.
Then, 
\begin{align}
 ( R \otimes W ) \; \bmu_0 &= 
 G_{\mu_0}  \operatorname{vec}(R) \ \ \mbox{and} \ \ 
 (R_0 \otimes W) \; \bmu_0 = G_{\mu_0} \operatorname{vec}(R_0).
\end{align}
Subtracting the two equations gives
\begin{align}
   \big\{ (R - R_0) \otimes W \big\} \bmu_0 - X (\bbeta_0 - \bbeta)
 = G_{\mu_0} \operatorname{vec}(R - R_0) - X (\bbeta_0 - \bbeta) = \0.
 \label{eq:r}
\end{align}
Since the coefficient matrix $\big[\  G_{\mu_0} , \ X \ \big]$ in Eq.~(24) has full column rank, the only solution is 
$\operatorname{vec}(R - R_0) = \0$ and $\bbeta - \bbeta_0 = \0$.
Hence, $ R = R_0 $ and $ \bbeta = \bbeta_0 $, 
which establishes the identifiability of $ (R,\, \bbeta) $.

The rank condition is satisfied whenever the explanatory variables and the spatially lagged mean vectors are linearly independent, 
excluding degenerate designs.
Although $ \bmu_0 $ is unknown, the proposition provides a sufficient population-level identification condition.

\smallskip 
{\bf Proof of Proposition 2.} \ \ 
From Eq.~(\ref{eqn:s_r}), the covariance matrix of $\bu = {S_\Lambda}^{-1} \bepsilon $ is given by
\begin{align}
 & V(\Lambda, \; \Sigma)
  = {S_\Lambda}^{-1} (\Sigma \otimes I_n) {S'_\Lambda}^{-1}.
\end{align}
Suppose that the equality $V(\Lambda, \; \Sigma) = V(\Lambda_0, \; \Sigma_0)$ holds.
This is equivalent to
$ V(\Lambda, \, \Sigma)^{-1} = V(\Lambda_0,\, \Sigma_0)^{-1} $.
This equality leads to ${S'}_\Lambda (\Sigma^{-1} \otimes I_n) {S_\Lambda} = {S'}_{\Lambda_0} ({\Sigma_0}^{-1} \otimes I_n) {S_{\Lambda_0}}$,
which is equivalent to
\begin{align}
   & \ \Sigma^{-1} \otimes I_n 
 - ( \Lambda' \Sigma^{-1} ) \otimes W'
 - ( \Sigma^{-1} \Lambda )  \otimes W
 + ( \Lambda' \Sigma^{-1} \Lambda ) \otimes ( W'W )  \\
 =
   & \ \Sigma_0^{-1} \otimes I_n 
 - ( \Lambda_0' \Sigma_0^{-1} ) \otimes W'
 - ( \Sigma_0^{-1} \Lambda_0 )  \otimes W
 + ( \Lambda_0' \Sigma^{-1}_0 \Lambda_0 ) \otimes ( W'W ).
 \notag
\end{align}
Since
$I_n$, $W$, $W'$, and $W'W$ are linearly independent,
each matrix coefficient must coincide separately.
Hence, 
$   \Sigma^{-1}                  = \Sigma_0^{-1}, \quad 
    \Lambda'\Sigma^{-1}          = \Lambda_0'\Sigma_0^{-1}, \quad 
    \Sigma^{-1}\Lambda           = \Sigma_0^{-1}\Lambda_0, \quad
    \Lambda' \Sigma^{-1} \Lambda = \Lambda_0' \Sigma^{-1}_0 \Lambda_0 
$. 
Since $ \Sigma = \Sigma_0, $ the equality $ \Sigma^{-1}\Lambda = \Sigma_0^{-1}\Lambda_0 $ 
immediately implies $ \Lambda = \Lambda_0. $
Hence, both $\Lambda$ and $\Sigma$ are identifiable.

\section*{Appendix B: Determinants of Kronecker-structured matrices}     

Let $\omega_i$ denote the eigenvalues of the spatial weight matrix $W$ for $i=1,\ldots,n$, 
and let $e_j(U)$ be the elementary symmetric polynomials of the eigenvalues of a matrix $U: K \times K$ for $j=1,\ldots,K$. 
Then, by the characteristic polynomial expansion, we have
\begin{equation}
\big| I_{Kn} - U \otimes W \big|
 = \prod_{i=1}^{n} \big| I_K - \omega_i U \big| \notag 
 = \prod_{i=1}^{n} 
    \left\{
      1 + \sum_{j=1}^{K} (-1)^j \, e_j(U)\,\omega_i^{\,j}
    \right\}.
\end{equation}
The elementary symmetric polynomials $e_j(U)$ can be expressed in terms of the power sums
$ \displaystyle
r_j = \mathrm{tr}(U^j), \ j=1,\ldots,K
$
through the Newton--Girard identities \citep{macdonald1995}:
\begin{align}
j \times e_j(U)
=
\sum_{\ell=1}^{j}(-1)^{\ell-1} \, e_{j-\ell}(U)\; r_\ell,
\qquad
e_0(U)=1.
\label{eq:newton-girard}
\end{align}
Hence, for any $K$, the determinant can be recursively computed from $r_j$.
For example,
\begin{equation}
e_1(U) = r_1, \ \ 
e_2(U) = \frac{r_1^2-r_2}{2}, \ \ 
e_3(U) = \frac{r_1^3-3r_1r_2+2r_3}{6}.
\end{equation}
Higher-order terms are obtained recursively from Eq.~(27).

\section*{Appendix C: Penalized likelihood and asymptotic approximation}
\label{app:penalized_likelihood}

We examine the estimator in Eq.~(\ref{eq:pen_estimator}) for a prespecified model, suppressing $m$ for brevity. 
Let $N = Kn$, with fixed $K$ and $n \to \infty$, and allow the tuning parameter to depend on $N$. 
The penalized score equation is
$U_N(\hat{\btheta}_{\gamma_N}) - \gamma_N \, D \, \hat{\btheta}_{\gamma_N} = \0$, where
$U_N(\btheta) = \partial \, \ell_N(\btheta) \, / \, \partial \, \btheta$.

Assume that the true parameter $\btheta_0$ is interior, the model is
identifiable, $S_R$, $S_\Lambda$, and their inverses are uniformly bounded,
and the spatial weight matrices have uniformly bounded row and column sums.
Also suppose that
$N^{-1} \, \mathcal I_N(\btheta_0) \ \xrightarrow{p} \ \mathcal J(\btheta_0)$ and
$N^{-1/2} \, U_N(\btheta_0) \ \xrightarrow{d} \ N\{ \, \0, \ \mathcal K(\btheta_0) \, \}$,
where $\mathcal J(\btheta_0)$ and $\mathcal K(\btheta_0)$ are finite and positive definite.
Taylor expansion of the penalized score gives
\begin{align}
   \sqrt N (\hat{\btheta}_{\gamma_N} - \btheta_0)
 = \left\{ \frac{\mathcal I_N(\btheta_0)}{N} + \frac{\gamma_N}{N} D \right\}^{-1}
   \left\{ \frac{U_N(\btheta_0)}{\sqrt N} - \frac{\gamma_N}{\sqrt N} \, D \, \btheta_0 \right\} + o_p(1).
 \label{eq:pen_expansion}
\end{align}
Thus, $\gamma_N / N \to 0$ preserves consistency. 
If $\gamma_N/\sqrt N\to0$, the first-order shrinkage bias vanishes and
\begin{align}
 \sqrt N (\hat{\btheta}_{\gamma_N} - \btheta_0)
 \ \xrightarrow{d} \ 
 N \! \left[ \, \0, \ \mathcal J(\btheta_0)^{-1} \, \mathcal K(\btheta_0) \, \mathcal J(\btheta_0)^{-1} \, \right].
 \label{eq:pen_asym_normal}
\end{align}
Under correct likelihood specification, the information identity reduces the
limiting covariance to $\mathcal J(\btheta_0)^{-1}$. 
If instead $\gamma_N / \sqrt N \to c > 0$, Eq.~(\ref{eq:pen_expansion}) has limiting mean
$- c \mathcal J(\btheta_0)^{-1} \, D \, \btheta_0$, explicitly showing the first-order shrinkage bias. 
Because the implementation searches a fixed bounded region for
$\gamma$, the negligible-penalty condition holds asymptotically for a prespecified model, 
although finite-sample bias may remain.

For fixed $(m,\gamma)$, linearization yields the covariance approximation in Eq.~(\ref{eq:pen_covariance}) 
and the effective degrees of freedom in Eq.~(\ref{eq:effective_df}). 
Neither expression includes the additional uncertainty from selecting $(m,\gamma)$. 
Hence, the expansion above does not establish the distribution conditional on pAIC or pBIC selection; 
bias correction and selection-adjusted inference for the MGNST model remain topics for future research.

\section*{Appendix D: Principal components of $\by_1,\ldots,\by_K$}   
\label{app:pca}
Consider the linear combination 
$\bz(\ba) = \sum_{k=1}^K a_k\by_{k} = (\ba' \otimes I_n) \by$,
where $\ba = (a_1,\ldots,a_K)'$ and $\|\ba\| = 1$. 
Writing $\by = \by_t : Kn \times 1$, let its conditional covariance matrix be
\[
 \Xi \ = \ \operatorname{Cov}(\, \by_t \, \mid \, \by_{t-1}, X_t \, )
     \ = \ S_R^{-1} \, S_\Lambda^{-1} (\Sigma \otimes I_n) \, S_\Lambda^{-T} \, S_R^{-T}
     \ = \ \big[\, \Xi_{k\ell}\, \big]_{k,\ell=1}^K : Kn \times Kn,
 \quad \Xi_{k\ell} : n \times n.
\]
Then
$\operatorname{Cov}\{\bz(\ba)\} = \sum_{k,\ell} \, a_k \, a_\ell \, \Xi_{k\ell}$, 
and its total variance over the $n$ spatial units is
\[
   \operatorname{tr} \left[ \, \operatorname{Cov}\{\bz(\ba)\} \, \right]
 = \ba' \, \widetilde{\Xi} \, \ba, 
 \ \ \mbox{where} \ \ 
   \widetilde{\Xi}
 = \left[ \ \operatorname{tr}(\, \Xi_{k\ell} \,) \ \right]_{k,\ell=1}^K : K \times K.
\]

Let
$\widetilde{\Xi} = M \, \Delta \, M'$,
where
$\Delta = \operatorname{diag}(\delta_1, \ldots, \delta_K)$,
$\delta_1 \geq \cdots \geq \delta_K \geq 0$, and $M = [\,\bmu_1,\ldots,\bmu_K\,]$ 
is an $K \times K$ orthogonal matrix with $ \bmu_r = (\mu_{1 r},\ldots,\mu_{K r})' $. 
The $r$-th principal component is
\[
   \mathrm{PC}_r
 = (\bmu_r'\otimes I_n) \, \by
 = \sum_{k=1}^K \, \mu_{kr}\by_k,
   \qquad
 \mathrm{PC}_{r,i} = \sum_{k=1}^K \, \mu_{kr} \, y_{k,i}.
\]
Thus, $\mathrm{PC}_1$ maximizes the total variance, which equals $\delta_1$, 
whereas $\mathrm{PC}_K$ has the minimum total variance $\delta_K$.
For $r\neq s$,
$\operatorname{tr}\{\operatorname{Cov}(\mathrm{PC}_r, \mathrm{PC}_s)\} = 0$;
hence, the components are orthogonal with respect to their spatially aggregated
cross-covariances, although they may remain correlated within and across spatial
units.


\begin{thebibliography}{99}

 \bibitem[Anselin(1988)]{anselin1988}
  Anselin, L. (1988).
  \textit{Spatial Econometrics: Methods and Models}.
  Kluwer Academic Publishers.

 \bibitem[Anselin(2003)]{anselin2003}
  Anselin, L. (2003). 
  ``Spatial econometrics," In \textit{A Companion to Theoretical Econometrics}, 
  B. H. Baltagi (ed.), Blackwell, 310--330. \ \ 
  doi:10.1002/9780470996249.ch15

 \bibitem[Banerjee, Carlin and Gelfand(2014)]{banerjee2014}
  Banerjee, S., Carlin, B. P. and Gelfand, A. E. (2014).
  \textit{Hierarchical Modeling and Analysis for Spatial Data} (2nd ed.).
  Chapman \& Hall/CRC, Boca Raton.

 \bibitem[Bradley, Holan and Wikle(2018)]{bradley2018}
  Bradley, J. R., Holan, S. H. and Wikle, C. K. (2018).
  ``Computationally efficient multivariate spatio-temporal models for high-dimensional count-valued data 
   (with Discussion)," 
  \textit{Bayesian Analysis}, \textbf{13}(1), 253--310. \ \ 
  doi:10.1214/17-BA1054

 \bibitem[Burridge et~al.(2017)]{burridge2017}
  Burridge, P., Elhorst, J. P., Zigova, K. and Sen, A. (2017).
  ``The SARAR model: a generalisation of the spatial autoregressive model,"
  \textit{Regional Science and Urban Economics}, \textbf{66}, 163--175. \ \ 
  doi:10.1016/j.regsciurbeco.2017.07.001
 
 \bibitem[Cressie and Wikle(2011)]{cressie2011}
  Cressie, N. and Wikle, C. K. (2011).
  \textit{Statistics for Spatio-Temporal Data}.
  Wiley, Hoboken.

 \bibitem[Elhorst(2014)]{elhorst2014}
  Elhorst, J. P. (2014).
  \textit{Spatial Econometrics: From Cross-Sectional Data to Spatial Panels}, Springer, London.
 
 \bibitem[Elliott and Wartenberg(2004)]{elliott2004}
 Elliott, P. and Wartenberg, D. (2004).
 ``Spatial epidemiology: Current approaches and future challenges,"
 \textit{Environmental Health Perspectives},
 \textbf{112}(9), {998--1006}.
 doi:{10.1289/ehp.6735}

 \bibitem[Halleck Vega and Elhorst(2016)]{halleckvega2016}
  Halleck Vega, S. and Elhorst, J. P. (2016).
  ``A regional unemployment model simultaneously accounting for serial dynamics, spatial dependence 
    and common factors,"
  \textit{Regional Science and Urban Economics},
  \textbf{60}, {85--95}.
  doi:10.1016/j.regsciurbeco.2016.07.002

 \bibitem[Hastie, Tibshirani and Friedman(2009)]{hastie2009}
 Hastie, T., Tibshirani, R. and Friedman, J. (2009).
 \textit{The Elements of Statistical Learning}.
 Springer.

 \bibitem[Huerta et al.(2004)]{huerta2004}
  Huerta, G., Sans{\'o}, B. and Stroud, J. R. (2004).
  ``A spatiotemporal model for Mexico City ozone levels,"
  \textit{Journal of the Royal Statistical Society: Series C (Applied Statistics)},
  \textbf{53} (2), 231--248.
  doi:10.1046/j.1467-9876.2003.05100.x

 \bibitem[Kelejian and Prucha(2004)]{kelejian2004}
 Kelejian, H. H. and Prucha, I. R. (2004).
 ``Estimation of simultaneous systems of spatially interrelated cross sectional equations,"
 \textit{Journal of Econometrics}, \textbf{118}, 27--50. 
 doi:10.1016/S0304-4076(03)00133-7

 \bibitem[Kilian and L{\"u}tkepohl(2017)Kilian and L{\"u}tkepohl]{kilian2017}
  Kilian, L. and L{\"u}tkepohl, H. (2017).
  \textit{Structural Vector Autoregressive Analysis}.
  Cambridge University Press, Cambridge.

 \bibitem[Konishi and Kitagawa(1996)]{konishi1996}
  Konishi, S. and Kitagawa, G. (1996).
  ``Generalised information criteria in model selection,''
  \textit{Biometrika}, \textbf{83}(4), 875--890.
  doi:10.1093/biomet/83.4.875

 \bibitem[Lee and Yu(2016)]{lee2016}
  Lee, L.-F. and Yu, J. (2016).
  ``Identification of spatial Durbin panel models," \textit{Journal of Applied Econometrics}, 
  \textbf{31}(1), 133--162.
  doi:10.1002/jae.2450
 
 \bibitem[LeSage and Pace(2009)]{lesage2009}
  LeSage, J. P. and Pace, R. K. (2009). 
  \textit{Introduction to Spatial Econometrics.} CRC Press.
 
 \bibitem[L{\"u}tkepohl(2005)]{lutkepohl2005}
  L{\"u}tkepohl, H. (2005).
  \textit{New Introduction to Multiple Time Series Analysis}.
  Springer, Berlin Heidelberg.

 \bibitem[Macdonald(1995)]{macdonald1995}
  Macdonald, I.~G. (1995).
  \textit{Symmetric Functions and Hall Polynomials} (2nd ed.).  Oxford University Press.

 \bibitem[Mastrantonio et~al.(2019)]{mastrantonio2019}
  Mastrantonio, G., Lasinio, G. J., Pollice, A., Capotorti, G., Teodonio, L., Genova, G. and Blasi, C. (2019).
  ``A hierarchical multivariate spatio-temporal model for clustered climate data with annual cycles," 
  \textit{The Annals of Applied Statistics}, \textbf{13}(2), 797--823. \ \ 
  doi:10.1214/18-AOAS1212
 
 \bibitem[Nishii, Ohta and Tanaka(2025)]{nishii2025}
  Nishii, R., Ohta, S. and Tanaka, S. (2025).
  Proposal and estimation of a simultaneous equation model for multiple spatio-temporal data.
  In \textit{Proceedings of the 2025 Japanese Joint Statistical Meeting}, p.~360. [In Japanese.]

 \bibitem[Nocedal and Wright(2006)]{nocedal2006}
  Nocedal, J. and Wright, S. J. (2006).
  \textit{Numerical Optimization}.
  Springer, New York.

 \bibitem[Ohta, Tanaka and Nishii(2026a)]{ohta2026a}
  Ohta, S., Tanaka, S. and Nishii, R. (2026a).
  MGNST Project Repository:
  Reproducible Data, Software and Numerical Results for Multivariate Spatio-Temporal Regression.
  GitHub repository.
  \url{https://github.com/s-ohta-s/mstr-RSS-2026}

 \bibitem[Ohta, Tanaka and Nishii(2026b)]{ohta2026b}
  Ohta, S., Tanaka, S. and Nishii, R. (2026b).
  Multivariate spatio-temporal modeling for regional GIS data:
  A statistical framework for analyzing multidimensional spatial interactions.
  \textit{The International Archives of the Photogrammetry, Remote Sensing and Spatial Information Sciences}, 
  forthcoming.

 \bibitem[Ward and Gleditsch(2019)]{ward2019}
  Ward, M. D. and Gleditsch, K. S. (2019). 
  \textit{Spatial Regression Models} (2nd ed.). SAGE Publications. 

 \bibitem[Yang and Lee(2017)]{yang2017}
  Yang, K. and Lee, L.-F. (2017).
  ``Identification and QML estimation of multivariate and simultaneous equations spatial autoregressive models,"
  \textit{Journal of Econometrics}, \textbf{196}(1), 196--214. 
  doi:10.1016/j.jeconom.2016.04.019

\end{thebibliography}
\end{document}